\documentclass[aps,prb,twocolumn,amsmath,amssymb,superscriptaddress,floatfix,eqsecnum]{revtex4-1}

\usepackage{amsfonts}
\usepackage{amsthm}
\usepackage{mathrsfs}
\usepackage{cases}
\usepackage{url}
\usepackage{framed}
\usepackage{cancel,soul}
\usepackage[normalem]{ulem}
\usepackage[dvipdfmx]{graphicx}
\usepackage{multirow}
\usepackage[pdftex]{color}
\usepackage{bm}
\usepackage{bbm}
\usepackage{slashed} 
\usepackage[breaklinks=true,colorlinks,citecolor=blue,linkcolor=blue,urlcolor=blue]{hyperref}
\usepackage{cases}
\usepackage{float}
\usepackage{url}
\usepackage{framed}

\def\=={\raisebox{0.35pt}{$\mathrm{:}$}\!\!=}
\def\df{\,\raisebox{0.95pt}{.}\hspace{-2.78pt}\raisebox{2.95pt}{.}\!\!=\,}
\def\dfr{\,=\!\!\raisebox{0.95pt}{.}\hspace{-2.78pt}\raisebox{2.95pt}{.}\,}

\newcommand{\bea}{\begin{eqnarray}}
\newcommand{\eea}{\end{eqnarray}}

\begin{document}

\title{Stability against contact interactions of a topological superconductor 
in two-dimensional space protected by time-reversal and
reflection symmetries} 
\author{\"{O}mer M. Aksoy}
\affiliation{Condensed Matter Theory Group, Paul Scherrer Institute, CH-5232 Villigen PSI, Switzerland}
\author{Jyong-Hao Chen}
\affiliation{Department of Physics, University of California, Berkeley, California 94720, USA}
\affiliation{Condensed Matter Theory Group, Paul Scherrer Institute, CH-5232 Villigen PSI, Switzerland}
\author{Shinsei Ryu}
\affiliation{Department of Physics, Princeton University, Princeton, New Jersey 08540, USA}
\author{Akira Furusaki}
\affiliation{Condensed Matter Theory Laboratory, RIKEN, Wako, Saitama 351-0198, Japan}
\affiliation{RIKEN Center for Emergent Matter Science (CEMS), Wako, Saitama 351-0198, Japan}
\author{Christopher Mudry}
\affiliation{Condensed Matter Theory Group, Paul Scherrer Institute, CH-5232 Villigen PSI, Switzerland}
\affiliation{Institut de Physique, EPF Lausanne, Lausanne CH-1015, Switzerland}
\date{\today}

\begin{abstract}
We study the stability of 
topological crystalline superconductors
in the symmetry class DIIIR and in two-dimensional space
when perturbed by quartic contact interactions.
It is known that no less than
eight copies of helical pairs of Majorana edge
modes can be gapped out by an appropriate interaction 
without spontaneously breaking any one of the protecting symmetries.
Hence, the noninteracting classification $\mathbb{Z}$ reduces to 
$\mathbb{Z}^{\,}_{8}$ when these interactions are present.
It is also known that
the stability when there are less than eight modes
can be understood in terms of the presence of topological
obstructions in 
the low-energy bosonic effective theories, which
prevent opening of a gap. Here, we investigate the 
stability of the edge theories with four, two, and one edge modes,
respectively.
We give an analytical derivation of the topological term for the 
first case, because of which the edge theory remains gapless. 
For two edge modes, we employ bosonization methods to derive an
effective bosonic action. When gapped, this bosonic theory
is necessarily associated to the
spontaneous symmetry breaking of either
one of time-reversal or reflection symmetry
whenever translation symmetry remains on the boundary.
For one edge mode, stability is explicitly established in the Majorana
representation of the edge theory.
\end{abstract}

\maketitle

\section{Introduction}

Topological phases of matter have attracted ever-growing
attention since the discovery of the integer quantum Hall
effect in 1980%
~\cite{klitzing1980new}.
One way of understanding such phases of matter 
is that, in the space of gapped Hamiltonians with certain symmetries,
there exist equivalence classes labeled by topological invariants.
Two Hamiltonians with distinct topological invariants are topologically 
inequivalent since they cannot be smoothly deformed into 
one another without a discontinuous change of the topological invariant.
In particular, a gap-closing phase transition must occur
under a parametric change between two topologically distinct phases
when the phase transition is continuous. If such
phases of matter are short-range entangled (as opposed to
long-range entangled), they are called 
symmetry protected topological (SPT) phases.
A central theme in the study of topological matter
is the classification of SPT phases.

SPT phases of noninteracting fermionic systems,
which are examples of fermionic SPT (FSPT) phases,
with nonspatial symmetries are
well understood and described by the tenfold way%
~\cite{PhysRevB.78.195125,Ryu_2010,kitaev2009}.
Such phases are classified according to the 
spatial dimension of the physical system and the 
absence or presence of local nonspatial symmetries,
namely, time-reversal (TR), particle-hole (PH), chiral,
and their combinations. 
The distinctive property of these phases is that
they are gapped with a nondegenerate ground state
when imposing periodic boundary conditions, 
while they support gapless boundary states on their boundaries
when imposing open boundary conditions
\footnote{
The notion of a spectral gap only applies if translation symmetry holds.
When translation symmetry is broken by a static random potential that
respects  the protecting symmetries and is weak relative to the spectral gap,
all eigenstates within the gap are insulating. The boundary states
that were gapless with translation symmetry along the boundary
remain delocalized in the presence of the static disorder potential.
		}.
This property is often caricatured by stating the simultaneous
existence of a gap in the bulk and of gapless boundary states.
  
A natural extension when classifying FSPT phases is the 
inclusion of spatial symmetries which are relevant to 
crystalline materials. 
It has been shown that spatial symmetries
such as reflection/inversion, point-group,
and space-group symmetries
can modify the topological classification or lead to distinct phases%
~\cite{PhysRevLett.106.106802,Ando2015,
PhysRevB.76.045302,PhysRevB.88.075142,PhysRevLett.111.056403,
PhysRevX.7.011020,PhysRevB.90.115207,PhysRevB.90.085304,
PhysRevB.88.125129,Slager2013,PhysRevB.87.035119,
PhysRevB.88.085110,PhysRevB.90.165114,PhysRevB.96.195109}. 
Moreover, the particular case of reflection
symmetry shows that the algebra between 
spatial and nonspatial symmetries also affects the classification%
~\cite{PhysRevB.88.075142,PhysRevB.88.125129,PhysRevB.90.165114,
PhysRevB.96.195109}.
There exist 27 symmetry classes with reflection symmetries as opposed 
to 10 Altland-Zirnbauer symmetry classes%
~\cite{PhysRevB.88.075142}.

Prior to the seminal papers in
Refs.\ \onlinecite{PhysRevB.81.134509,PhysRevB.83.075103,kitaev2009},
it was believed that the FSPT classification was robust to
symmetry-preserving local interactions. 
However, Fidkowski and Kitaev showed that
the noninteracting boundary theory with eight zero modes
for the symmetry class BDI in one-dimensional space is unstable to
local and symmetry-preserving interactions by constructing
such an interaction and showing that the interacting theory is 
adiabatically connected to a gapped boundary theory
without any spontaneous breaking of the protecting symmetry. 
Hence, the noninteracting classification by group $\mathbb{Z}$
reduces to the classification by group $\mathbb{Z}^{\,}_{8}$
in the presence of local interactions compatible with
the BDI symmetry class in one-dimensional space.
Following this work, effects of interactions have also 
been an important extension of the classification of
noninteracting FSPT phases when
local symmetry preserving interactions are present.
These effects have been investigated in Refs.\ 
\onlinecite{PhysRevB.88.064507,2012PhRvB..85x5132R,
Qi_2013,PhysRevX.3.041016,
metlitski2014interaction,
PhysRevB.89.195124,
PhysRevB.92.125104,PhysRevB.90.245120,
Kapustin2015,Wang629,PhysRevLett.109.096403,
PhysRevLett.117.206405,freed2016reflection,
PhysRevB.95.195108}.

In particular, Morimoto {\it et al}.\ in Ref.\ \onlinecite{PhysRevB.92.125104} 
illustrated how the topological classification
for all tenfold symmetry classes and spatial dimensions 
can change by symmetry-preserving
quartic contact interactions. They found that
the noninteracting classification in terms of group
$\mathbb{Z}^{\,}_{2}$
is always stable to symmetry-preserving quartic contact interactions,
while that with group $\mathbb{Z}$ is only
stable in even spatial dimensions. 
For symmetry classes with the $\mathbb{Z}$ invariant in odd space dimensions,
there is a breakdown to $\mathbb{Z}^{\,}_{N}$ with
some positive integer $N$. The method employed to show 
the breakdown relies on the presence or absence of 
topological obstructions in the low-energy bosonic theories describing
$\nu$ copies of boundary states. These bosonic theories arise when one 
decouples the quartic interactions with a Hubbard-Stratonovich 
transformation by introducing dynamical (mass) fields 
and one integrates over the fermionic degrees of freedom
in the path-integral description of quantum mechanics.
The low-energy sector is then described by a nonlinear sigma model
(NLSM) whose target space might support topological obstructions. 
When a local topological obstruction is present,
it is conjectured in Ref.\ \onlinecite{PhysRevB.92.125104}
that the ground state in the thermodynamic limit 
either remains gapless or becomes gapped  
with spontaneous (discrete) symmetry breaking.
\footnote{
In other words,
any local topological obstruction implies either
a gapless ground state or gapped but degenerate ground states
in the thermodynamic limit. Conversely,
a nondegenerate ground state in the thermodynamic limit implies
no local topological obstruction. However, the absence of any
local topological obstruction does not imply 
a nondegenerate ground state in the thermodynamic limit.
		}
A topological obstruction is present whenever 
at least one of the homotopy groups 
$\pi^{\,}_{l}(\mathsf{S}^{\mathrm{N}(\nu)-1})$
is nontrivial%
~\cite{abanov2000theta}, 
where $l=0,\cdots,d+1$, $d$ is the spatial
dimension of the system, $\mathsf{S}^{\mathrm{N}(\nu)-1}$ is the
$\left(\mathrm{N}(\nu)-1\right)$-dimensional sphere, and $\mathrm{N}(\nu)$ 
is the maximum number of pairwise anticommuting dynamical mass 
matrices for the $\nu$ boundary modes
(the upper bound on $l$ is here dictated by locality).
The number $\mathrm{N}(\nu)$
grows with the number of boundary modes $\nu$ (with plateaus
when $2^{n-1}<\nu<2^{n}$ for some integer $n$). 
Once the inequality $\mathrm{N}(\nu)-1>d+1$ is satisfied, all
homotopy groups $\pi^{\,}_{l}(\mathsf{S}^{\mathrm{N}(\nu)-1})$ with
$l=0,\cdots,d+1$ vanish so that no topological obstructions compatible
with locality are present.	
The smallest number of boundary modes
$\nu^{\,}_{\mathrm{min}}$
for which locality prevents the presence of a topological obstruction
is conjectured in Ref.\ \onlinecite{PhysRevB.92.125104} to
determine the breakdown pattern, i.e.,
the $\mathbb{Z}$ classification is reduced to the cyclic group
$\mathbb{Z}\hbox{ mod $\nu^{\,}_{\mathrm{min}}$ }=
\mathbb{Z}^{\,}_{\nu^{\,}_{\mathrm{min}}}$.

This strategy relying on the presence or absence of
topological obstructions 
in NLSMs to study the effects of interactions on
FSPT phases has also been applied to the crystalline topological
phases. Song and Schnyder in Ref.\
\onlinecite{PhysRevB.95.195108} 
have shown the patterns for the breakdown of the noninteracting
FSPT classification for all  crystalline symmetry classes with reflection or
twofold rotation symmetries by quartic contact interactions.

In this paper, we focus on TR symmetric
two-dimensional crystalline topological superconductors.
They realize the FSPT phase DIII.
In addition to time-reversal symmetry (TRS), we impose reflection symmetry (RS)
and denote with DIIIR a superconducting phase
protected by TRS, RS, and translation symmetry (TS).
The noninteracting FSPT of symmetry class DIII in two spatial dimensions has
a $\mathbb{Z}^{\,}_{2}$ group structure.
The noninteracting FSPT of symmetry class DIIIR in two spatial dimensions has
a $\mathbb{Z}$ group structure
when reflection has a unitary Hermitian representation that
anticommutes with the representations of both the reversal of time
and the interchange of particles and holes
\cite{PhysRevB.88.064507,PhysRevB.88.075142}\textsuperscript{,}
\footnote{Had we chosen a unitary Hermitian representation of reflection
that realizes a different algebra with the time-reversal or 
the interchange of particles and holes, the classification 
would be either trivial or have a $\mathbb{Z}^{\,}_{2}$ group structure.}.

The interacting classification of the FSPT symmetry class
DIIIR in two-dimensional space was first obtained in Ref.\
\onlinecite{PhysRevB.88.064507}
by showing that no less than eight copies of the gapless edge theory 
can be gapped without spontaneous breaking of the protecting
symmetries by local interactions.
Hence, local and symmetry-preserving interactions
reduce the noninteracting classification
with group $\mathbb{Z}$
to the one with group $\mathbb{Z}^{\,}_{8}$. 
The same result was later obtained in Refs.\
\onlinecite{PhysRevB.92.125104,PhysRevB.95.195108} by showing
on general grounds the possibility for the
existence of topological obstructions
in the NLSM description of 
the edge theory using homotopy arguments.
However, it remains open to (i) \textit{explicitly} construct these
topological obstructions in the low-energy theory
and to (ii) \textit{explicitly} show how the gapless edge modes
remain stable in the presence of local and symmetry-preserving interactions
because of the topological obstructions.

We start from an FSPT symmetry class DIIIR in two spatial dimensions
and aim to derive its low-energy bosonic action
along one of its connected boundaries explicitly.
The fermionic edge theory is described 
by $\nu$ copies of helical Majorana fields.
If interactions are absent, this edge theory cannot be gapped,
whatever the value of $\nu\in\mathbb{Z}$.
If quartic contact interactions are present,
Table \ref{tab: 2D DIIIR--}
implies the existence of a topological obstruction
when $\nu=1$, another one when $\nu=2,3$, and the last one when $\nu=4,5,6,7$.
The goal of this paper is to construct explicitly
these topological obstructions. To achieve this end,
it is sufficient to
start from an interacting boundary Hamiltonian for 
$\nu=1$, $\nu=2$, and $\nu=4$ copies
of relativistic helical Majorana fields, 
respectively\,\cite{PhysRevB.92.125104}.

There are three possibilities when local interactions
that preserve the TRS, RS and TS are included:
(i) the edge theory remains gapless,
(ii) the edge theory is gapped but some of the protecting 
symmetries are spontaneously broken, and
(iii) the edge theory is gapped without any symmetry breaking.
We call the edge theory unstable against the given interactions
when the last scenario is realized, otherwise we
say that it is stable.

\begin{table}[tb]
\caption{
Reduction from $\mathbb{Z}$ 
to
$\mathbb{Z}^{\,}_{8}$ due to interactions
for the topologically equivalent classes of the
two-dimensional topological superconductors protected by
time-reversal, reflection, and translation symmetries (DIIIR).
If $V^{\,}_{\nu}$ denotes the space of
$\nu\times\nu$ normalized dynamical Dirac mass matrices on
a boundary invariant under both reflection and translation, then
the limit $\nu\to\infty$ of these spaces is the classifying space
$R^{\,}_{1}$ \cite{PhysRevB.92.125104}.  
The second column shows the stable $D$-th homotopy groups 
of the classifying space $R^{\,}_{1}$. 
The third column gives the number $\nu$ of copies of 
boundary (Dirac) fermions for which a topological obstruction is permissible.
The fourth column gives the type of topological obstruction
that prevents the gapping of the boundary (Dirac) fermions. 
We note that for $D=7$, even though the $\pi^{\,}_{7}(R^{\,}_{1})$ is
nonvanishing there is no topological obstruction that is
compatible with locality. The topological obstruction 
for $\nu=2$ is also present when $\nu=3$
and the topological  obstruction for $\nu=4$ is also present when
$\nu=5,6,7$.
\label{tab: 2D DIIIR--}
}
\begin{tabular}{ccccccc}
\hline \hline
$D$ 
&\qquad\qquad&
$\pi^{\,}_{D}(R^{\,}_{1})$ 
&\qquad\qquad&
$\nu$  
&\qquad\qquad& 
Topological obstruction 
\\
\hline
0&&$\mathbb{Z}^{\,}_{2}$ && 1  && Domain wall  \\
1&&$\mathbb{Z}^{\,}_{2}$ && 2  && Vortex       \\
2&&0                   &&    &&              \\
3&&$\mathbb{Z}$        && 4  &&  WZ term     \\
4&&0                   &&    &&              \\
5&&0                   &&    &&              \\
6&&0                   &&    &&              \\
7&&$\mathbb{Z}$        && 8  &&  None        \\
\hline \hline
\end{tabular}
\end{table}

Out of three cases ($\nu=1,2,4$) considered, 
we derive the topological obstruction 
(i.e., the topological term in the NLSM)
for $\nu=4$ case explicitly via the gradient expansion
of a fermionic determinant.
We identify $\mathrm{N}(4)=4$ possible quartic interactions
which leads to the nontrivial homotopy
group $\pi^{\,}_{3}(\mathsf{S}^{3})=\mathbb{Z}$.
The corresponding term is of the Wess-Zumino (WZ) type,
as is indicated by line $D=4$ and column 4
from Table \ref{tab: 2D DIIIR--}.
It is expected that NLSM action supplemented with
an appropriate WZ term is equivalent 
to an action for fermions in (1+1) dimensions at criticality\,
\cite{witten1984}. Hence, the edge theory remains stable.
 
For the remaining two cases,  
the method of gradient expansion turns out
to be difficult to apply. We identify
$\mathrm{N}(2)=2$ and $\mathrm{N}(1)=1$ interaction channels
for two and one edge modes, respectively.
In the former case, $\pi^{\,}_{1}(\mathsf{S}^{1})=\mathbb{Z}$
implies the existence of vortex configurations
of the NLSM bosonic field,
as indicated by line $D=1$ and column 4 from Table \ref{tab: 2D DIIIR--}.
In the latter case, $\pi^{\,}_{0}(\mathsf{S}^{0})=\mathbb{Z}^{\,}_{2}$
implies the existence of domain-wall configurations,
as indicated by line $D=0$ and column 4
from Table \ref{tab: 2D DIIIR--}. However, 
vortex configurations invalidate the gradient
expansion by being singular at the vortex core
and domain-wall configurations necessarily bind 
zero energy modes, which leads to vanishing 
fermionic determinants.

For the $\nu=2$ case, we follow an alternative approach
by adopting bosonization techniques. 
The low-energy edge theory turns out to be 
described by a sine-Gordon action. Such a 
theory is not necessarily gapless; however, we demonstrate
that any twofold degenerate and translation-invariant
gapped ground state spontaneously breaks either the
TRS or the RS.
This analysis is then supplemented
by considering interactions that, prior to bosonization,
are local momonials
in certain fermionic bilinears of arbitrary order
(i.e., not necessarily of quadratic order).
We find that spontaneous symmetry breaking
of either the TRS or RS
always occurs when the interaction is strong and of even order,
while these protecting symmetries are not broken
but TS is explicitly broken when the interaction is
strong and of odd order. In the latter case, there are no protected
delocalized edge states within the bulk gap, but there are ``corner'' states
on the boundary within the bulk gap.

For the case of $\nu=1$,
we consider smooth interpolations between certain mass profiles
and show
the existence of two topological sectors distinguished by
a $\mathbb{Z}^{\,}_{2}$ invariant. The existence of these
two topological sectors is closely related to a global
$\mathbb{Z}^{\,}_{2}$ anomaly.

The paper is organized as follows. 
In Sec. \ref{sec: def & sym}, we define the 
boundary Hamiltonian for $\nu$-edge modes
and the symmetries of the class DIIIR.
The subsequent sections discusses
the cases $\nu=4,2,1$, respectively.
We conclude with
Sec.\ \ref{sec:Conclusion}.

\section{Definitions and Symmetries}
\label{sec: def & sym}

We describe the one-dimensional boundary 
of a two-dimensional crystalline topological 
superconductor with the Hamiltonian
\begin{subequations}
\label{eq:Begining Hamiltonian defs}
\begin{align}
&\widehat{H}^{\,}_{\mathrm{bd}}
\,
\df 
\widehat{H}^{\,}_{0}
+
\widehat{H}^{\,}_{\mathrm{int}},
\label{eq:Begining Hamiltonian defs a}
\\
&
\label{eq:Begining Hamiltonian 0 defs}
\widehat{H}^{\,}_{0}
\,\,\,\,
\equiv
\int
\mathrm{d}x\,
\hat{\chi}^{\dag}
\mathcal{H}^{\,}_{0}
\hat{\chi}
\nonumber\\
&\qquad
\df
\int
\mathrm{d}x\,
\hat{\chi}^{\dag}
\left(
\sigma^{\,}_{3}\otimes\mathbbm{1}^{\,}_{\nu}\,\mathrm{i}\partial^{\,}_{x}
\right)
\hat{\chi},
\\
&
\widehat{H}^{\,}_{\mathrm{int}}
\,
\df
-
\int
\mathrm{d}x\,
\lambda^{2}
\sum_{l=1}^{\mathrm{N}(\nu)}
\left(
\hat{\chi}^{\dag}\,\beta^{\,}_{l}\,\hat{\chi}
\right)^{2}.
\end{align}
\end{subequations}
The Hamiltonian $\widehat{H}^{\,}_{0}$
describes  $\nu$ pairs of left- and right-moving quantum Majorana fields.
The components $\hat{\chi}^{\,}_{a}$
and $\hat{\chi}^{\dag}_{a}$ with $a=1,\cdots,2\nu$
of the quantum-fields and their adjoints obey the equal-time algebra
\begin{subequations}
\begin{align}
\left\{
\hat{\chi}^{\,}_{a}(x),
\hat{\chi}^{\,}_{a'}(x')
\right\}=
\delta^{\,}_{aa'}\,
\delta(x-x'),
\label{eq:majorana anticommutator defs}
\end{align}
with all other anticommutators vanishing
and we impose the Majorana condition 
\begin{equation}
\hat{\chi}^{\dag}=\hat{\chi}^{\mathsf{T}}. 
\label{eq:majorana condition defs}
\end{equation}
\end{subequations}
The Hamiltonian $\widehat{H}^{\,}_{\mathrm{int}}$ encodes the quartic contact 
interactions with coupling constant $\lambda^{2}$ between the $\nu$
different flavors. The matrix $\mathbbm{1}^{\,}_{\nu}$ is the
identity matrix in flavor space. The label
$l=1,\cdots,\mathrm{N}(\nu)$
enumerates all $2\nu\times2\nu$ Hermitian matrices such that
(i) they square to the identity
$\beta^{2}_{l}=\mathbbm{1}^{\,}_{2\nu}$,
(ii) any pair $(\beta^{\,}_{l},\beta^{\,}_{l'})$ anticommutes
pairwise as well as with $\sigma^{\,}_{3}\otimes\mathbbm{1}^{\,}_{\nu}$,
and (iii) each $\beta^{\,}_{l}$ is odd under 
complex conjugation.
The first two conditions restrict the $\mathrm{N}(\nu)$
interaction channels to the squares of
bilinears that are not competing Dirac mass terms.
The last condition follows from imposing 
a Majorana condition on the fermionic quantum fields 
as we do now. Had we demanded instead of (iii) that
each $\beta^{\,}_{l}$ is even under complex conjugation,
the bilinear $\hat{\chi}^{\dag}\,\beta^{\,}_{l}\,\hat{\chi}$
would then vanish because of the Majorana condition.
We emphasize that $\mathrm{N}(\nu)$ 
is constant when $2^{n-1}<\nu<2^{n}$ for some integer $n$.
This means that the target space corresponding to the normalized
dynamical Dirac masses does not change when $2^{n-1}<\nu<2^{n}$
for some integer $n$.

Following Refs.\ \onlinecite{PhysRevB.88.064507,PhysRevB.88.075142},
we define the PH, TR, and
reflection transformations,
\begin{subequations}
\label{eq:SymmetryOperators defs}
\begin{align}
\label{eq:SymmetryOperators a defs}
&\mathcal{C}^{\,}_{\mathrm{bd},\nu}\df
\mathbbm{1}^{\,}_{2}\otimes\mathbbm{1}^{\,}_{\nu}\,\mathsf{K},
\\
\label{eq:SymmetryOperators b defs}
&\mathcal{T}^{\,}_{\mathrm{bd},\nu}\df
\mathrm{i}\sigma^{\,}_{2}\otimes\mathbbm{1}^{\,}_{\nu}\,\mathsf{K},
\\
\label{eq:SymmetryOperators c defs}
&\mathcal{R}^{\,}_{\mathrm{bd},\nu}\df
\mathrm{i}\sigma^{\,}_{2}\otimes\mathbbm{1}^{\,}_{\nu},
\end{align}
\end{subequations}
where $\bm{\sigma}$ are the Pauli matrices and $\mathsf{K}$ denotes
complex conjugation. 
They satisfy the defining conditions 
of the symmetry class DIIIR, i.e.,
\begin{equation}
\mathcal{C}^{2}_{\mathrm{bd},\nu}=+1,
\qquad
\mathcal{T}^{2}_{\mathrm{bd},\nu}=-1,
\qquad
\mathcal{R}^{2}_{\mathrm{bd},\nu}=-1,
\end{equation}
with the algebra
\begin{equation}
\left[
\mathcal{C}^{\,}_{\mathrm{bd},\nu},
\mathcal{R}^{\,}_{\mathrm{bd},\nu}
\right]=0,
\qquad
\left[
\mathcal{T}^{\,}_{\mathrm{bd},\nu},
\mathcal{R}^{\,}_{\mathrm{bd},\nu}
\right]=0.
\end{equation}
Had we chosen the Hermitian representation 
for the reflection transformation \eqref{eq:SymmetryOperators c defs}, 
i.e., $\mathcal{R}^{\,}_{\mathrm{bd},\nu}=
\sigma^{\,}_{2}\otimes\mathbbm{1}^{\,}_{\nu}$,
it would anticommute with both PH and TR transformations.
This is consistent with the definition of DIIIR in
Ref.\ \onlinecite{PhysRevB.88.075142}.
The anti-Hermitian representation
\eqref{eq:SymmetryOperators c defs}
is chosen since the transformation law is then covariant
with respect to the Majorana condition 
\eqref{eq:majorana condition defs}.
Moreover, we demand that transformations \eqref{eq:SymmetryOperators defs}
are (spectral) symmetries of the single-particle Hamiltonian 
\eqref{eq:Begining Hamiltonian 0 defs}
\begin{subequations}
\label{eq:imposing DIIR symmetries defs}
\begin{align}
&
\mathcal{C}^{\,}_{\mathrm{bd},\nu}\,
\mathcal{H}^{\,}_{0}(x)
\,\mathcal{C}^{-1}_{\mathrm{bd},\nu}=
-\mathcal{H}^{\,}_{0}(x),
\label{eq:imposing DIIR symmetries defs a}
\\
&
\mathcal{T}^{\,}_{\mathrm{bd},\nu}\,
\mathcal{H}^{\,}_{0}(x)\,
\mathcal{T}^{-1}_{\mathrm{bd},\nu}=
+\mathcal{H}^{\,}_{0}(x),
\label{eq:imposing DIIR symmetries defs b}
\\
&
\mathcal{R}^{\,}_{\mathrm{bd},\nu}\,
\mathcal{H}^{\,}_{0}(-x)\,
\mathcal{R}^{-1}_{\mathrm{bd},\nu}=
+\mathcal{H}^{\,}_{0}(x).
\label{eq:imposing DIIR symmetries defs c}
\end{align}
\end{subequations}
When the conditions 
\eqref{eq:imposing DIIR symmetries defs}
are satisfied and we impose invariance under TS,
\begin{equation}
\widehat{T}(x')\,
\widehat{H}^{\,}_{\mathrm{bd}}\,
\widehat{T}^{-1}(x')
=
\widehat{H}^{\,}_{\mathrm{bd}},
\qquad\forall\,
x'\in\mathbb{R},
\label{eq:imposing DIIR symmetries defs d}
\end{equation}
where $\widehat{T}(x')$ is the operator
that implements the translation by $x'$,
then Hamiltonian
\eqref{eq:Begining Hamiltonian 0 defs}
cannot  be gapped by adding bilinears of the
fermionic fields
for any $\nu=1,2,3,\cdots$%
~\cite{PhysRevB.88.075142}.
In this case, the noninteracting classification for the class DIIIR is
$\mathbb{Z}$. However, bilinears that are odd under reflection are
allowed if they are multiplied by a (smooth) function of $x$,
a space-dependent mass, that is odd under $x\to-x$
and must thus vanish at the origin $x=0$.
Such a mass gaps the single-particle spectrum except
for a mid-gap bound state whose envelope decays exponentially fast
away from $x=0$. Any such mass breaks TS
and the origin can be thought of as a point defect or a ``corner''
along the one-dimensional boundary at which the mass term must change sign
if it is to respect reflection symmetry. In the presence of such 
a space-dependent mass, the noninteracting classification reduces to that 
of the symmetry class DIII, i.e., $\mathbb{Z}^{\,}_{2}$%
~\cite{PhysRevB.91.235111,PhysRevB.88.064507,PhysRevB.88.075142}.

Alternatively,
we can write down the partition function
\begin{subequations}
\label{eq:PartitionBeforeHS defs}
\begin{align}
&
Z^{\,}_{\mathrm{bd}}\df
\int\mathcal{D}[\chi]\,
e^{-S^{\,}_{\mathrm{bd}}},
\\
&
S^{\,}_{\mathrm{bd}}\df
\int\mathrm{d}\tau\mathrm{d}x\,
\mathcal{L}^{\,}_{\mathrm{bd}},
\\
&
\mathcal{L}^{\,}_{\mathrm{bd}}\df
\chi^{\dag}
\left(
\partial^{\,}_{\tau}
+
\sigma^{\,}_{3}\otimes\mathbbm{1}^{\,}_{\nu}\,
\mathrm{i}\partial^{\,}_{x}
\right)
\chi
-
\lambda^{2}
\sum_{l=1}^{\mathrm{N}(\nu)}
\left(
\chi^{\dag}\,\beta^{\,}_{l}\,\chi
\right)^{2},
\end{align}
where the action is defined on (1+1)-dimensional 
Euclidean space-time. The integration
variables are the components of the
 Grassmann-valued spinor $\chi$,
as $\chi^{\dag}$ is linearly constrained to $\chi$ through
the Majorana condition \eqref{eq:majorana condition defs}.
\end{subequations}
The interaction terms can be decoupled via 
Hubbard-Stratonovich transformation. The partition 
function \eqref{eq:PartitionBeforeHS defs}
then takes the form
\begin{subequations}
\label{eq:PartitionBeforeHS bis defs}
\small
\begin{align}
&
Z^{\,}_{\mathrm{bd}}=
\mathrm{const}\times
\int\mathcal{D}[\chi]\int\mathcal{D}[\phi^{\,}_{\beta^{\,}_{l}}]\,
e^{-S'^{\,}_{\mathrm{bd}}},
\\
&
S^{\prime}_{\mathrm{bd}}=
\int\mathrm{d}\tau\mathrm{d}x\,
\mathcal{L}^{\prime}_{\mathrm{bd}},
\\
&
\mathcal{L}^{\prime}_{\mathrm{bd}}=
\chi^{\dag}
\left(
\partial^{\,}_{\tau}+\mathcal{H}^{(\mathrm{dyn})}_{\mathrm{bd}}
\right)
\chi
+
\frac{1}{(2\lambda)^{2}}
\sum_{l=1}^{\mathrm{N}(\nu)}
\phi^{2}_{l},
\\
&
\mathcal{H}^{(\mathrm{dyn})}_{\mathrm{bd}}
\df
+
\sigma^{\,}_{3}\otimes\mathbbm{1}^{\,}_{\nu}\,
\mathrm{i}\partial^{\,}_{x}
+
\sum_{l=1}^{\mathrm{N}(\nu)}
\beta^{\,}_{l}\,\phi^{\,}_{l}.
\end{align}
\end{subequations}
We have thereby defined the dynamical single-particle boundary Hamiltonian
$\mathcal{H}^{(\mathrm{dyn})}_{\mathrm{bd}}$. 
Conditions \eqref{eq:imposing DIIR symmetries defs} on
$\mathcal{H}^{(\mathrm{dyn})}_{\mathrm{bd}}$ can be met as follows.
PHS imposes that
\begin{equation}
\mathsf{K}\,\beta^{\,}_{l}\,\mathsf{K}^{-1}=\beta^{*}_{l}=
-\beta^{\,}_{l},
\qquad
l=1,\cdots,\mathrm{N}(\nu),
\label{eq:PHS on beta defs}
\end{equation}
for any $\beta^{\,}_{l}$ Hermitian $2\nu\times2\nu$ matrix.
Hence, imposing the Majorana condition trivially satisfies
the PHS.
Once the maximum number of $\beta^{\,}_{l}$ matrices that are
compatible with PHS is found, the symmetry requirements coming from
TRS and RS can be satisfied by imposing
that $\phi^{\,}_{l}$ is
either odd or even under time-reversal and reflection. 
From now on, we shall use the shorthand notation for the $4^{n}$,
$2^{n}\times2^{n}$ Hermitian matrices
\begin{align}
&
\mathrm{X}^{\,}_{\mu^{\,}_{1} \mu^{\,}_{2} ... \mu^{\,}_{n}}\df
\sigma^{(1)}_{\mu^{\,}_{1}}
\otimes
\sigma^{(2)}_{\mu^{\,}_{2}}
\otimes
\sigma^{(3)}_{\mu^{\,}_{3}}
\otimes
\cdots
\otimes
\sigma^{(n)}_{\mu^{\,}_{n}},
\nonumber
\\
&
\left(
\mathrm{X}^{\,}_{\mu^{\,}_{1} \mu^{\,}_{2} ... \mu^{\,}_{n}}	
\right)^{2}
=
\mathbbm{1}^{\,}_{2^{n}},
\qquad
\mu^{\,}_{j}=\mathrm{0,1,2,3},
\label{eq:ShorthandMatrix defs}
\end{align}
where $\sigma^{(j)}_{0}$ is $\mathbbm{1}^{\,}_{2},$
$\bm{\sigma}^{(j)}$
are the associated Pauli matrices, and $n\in\mathbb{Z}$ 
is related to $\nu$ by the relation $2^{n-1}=\nu.$

The partition function \eqref{eq:PartitionBeforeHS bis defs}
is quadratic in Grassmann variables, 
which therefore can be integrated out
to yield an effective action of bosonic fields 
$\phi^{\,}_{\beta^{\,}_{l}}$,
provided the Majorana Pfaffian is nonvanishing.
This effective theory is described by the partition function
\begin{align}
\label{eq:effective part func defs}
Z
=
\int
\mathcal{D}[\bm{\phi}]\,
\delta\left(\bm{\phi}^{2}-\bar{\phi}^{2}\right)
e^{
-
\int
\mathrm{d}^{2}x\,
\frac{1}{g}
\left(\partial^{\,}_{\mu}\bm{\phi}\right)^{2}
+
\Gamma[\bm{\phi}]
}
\end{align}
where $\bar{\phi}^{2}>0$ is a  real-valued constant,
$\bm{\phi}$ is a $\mathrm{N}(\nu)$-dimensional vector
field that is normalized through the nonlinear 
constraint imposed by the $\delta$ function,
 and the symbol $\Gamma[\bm{\phi}]$ 
signifies the existence of a topological obstruction.
In other words, the presence of the symbol $\Gamma[\bm{\phi}]$
implies that the effective action 
associated to the partition function 
\eqref{eq:effective part func defs}
is not merely that of a NLSM.
Due to the nonlinear constraint imposed
on $\mathrm{N}(\nu)$ bosonic fields, the
target space in Eq.\ \eqref{eq:effective part func defs}
is the unit sphere $\mathsf{S}^{\mathrm{N}(\nu)-1}$.
The symbol $\Gamma[\bm{\phi}]$ is present in 
Eq.\ \eqref{eq:effective part func defs} whenever 
one of the homotopy groups,
\begin{align}
&\pi^{\,}_{0}(\mathsf{S}^{\mathrm{N}(\nu)-1}),\nonumber\\
&\qquad\pi^{\,}_{1}(\mathsf{S}^{\mathrm{N}(\nu)-1}),\nonumber\\
&\qquad\qquad \pi^{\,}_{2}(\mathsf{S}^{\mathrm{N}(\nu)-1}),\label{eq:homotopy groups}\\
&\qquad\qquad\qquad\cdots \nonumber\\
&\qquad\qquad\qquad\qquad\pi^{\,}_{d+1}(\mathsf{S}^{\mathrm{N}(\nu)-1}),\nonumber
\end{align}
is nontrivial \cite{abanov2000theta}. 
(The upper bound $d+1$ is imposed as 
topological obstructions corresponding to higher
homotopy groups modify the equations of motions in a nonlocal way
\cite{PhysRevB.92.125104}.)
Such topological obstructions are expected
to prevent gapping out the edge modes.
If no such topological obstructions exists, the low-energy
effective theory is described by no more than a NLSM action.
For space-time dimension two, the action then flows to 
the strong coupling $g\to\infty$ stable fixed point
\cite{POLYAKOV197579}.
This is the quantum disordered phase
that describes a gapped phase of matter that is symmetric
under all protecting symmetries.
The original $\nu$ gapless edge modes have been gapped by the interactions
without any of the preserving symmetries being spontaneously broken.
Hence, the noninteracting gapless edge theory is smoothly connected to
a strongly interacting gapped edge theory upon switching on local symmetry
preserving interactions.
The presence of the topological obstruction
manifests itself
by modifying the renormalization group (RG) flow 
and preventing the flow to the 
strong coupling limit $g\to\infty$.  

\section{The case $\nu=4$}
\label{sec:nu 4}

The set (\ref{eq:ShorthandMatrix defs}) with $n=3$ has the 64 elements
$\{\mathrm{X}^{\,}_{\mu\rho\sigma}\}$ with $\mu,\rho,\sigma=0,\cdots,3$.
This set
spans the space of $8\times8$ Hermitian matrices.
For $\nu=4$, there are at most $\mathrm{N}(4)=4$ interaction channels 
allowed by the symmetry conditions \eqref{eq:imposing DIIR symmetries defs},
each of which is labeled by the Hermitian $8\times8$ matrix $\beta^{\,}_{l}$.
We consider the parametrization
\begin{subequations}
\label{eq:dyn Ham nu 4}
\begin{equation}
\mathcal{H}^{\mathrm{dyn}}_{\mathrm{bd}}(\tau,x)\df
\beta^{\,}_{0}\,\mathrm{i}\partial^{\,}_{x}
+
\sum_{l = 1}^{4}
\beta^{\,}_{l}\,
\phi^{\,}_{l}(\tau,x)
\end{equation}
of the dynamical boundary single-particle Hamiltonian,
where without loss of generality, we make the choice
\begin{align}
&
\beta^{\,}_{0} \df \mathrm{X}^{\,}_{300},
\\
&
\beta^{\,}_{1} \df \mathrm{X}^{\,}_{210},
\\
&
\beta^{\,}_{2} \df \mathrm{X}^{\,}_{230},
\\
&
\beta^{\,}_{3} \df \mathrm{X}^{\,}_{222},
\\
&
\beta^{\,}_{4} \df \mathrm{X}^{\,}_{102}=
-
\mathrm{X}^{\,}_{300}\,
\mathrm{X}^{\,}_{210}\,
\mathrm{X}^{\,}_{230}\,
\mathrm{X}^{\,}_{222}.  
\end{align}
The choice
$\{\beta^{\,}_{1},\beta^{\,}_{2},\beta^{\,}_{3},\beta^{\,}_{4}\}$
is not unique,
but this lack of uniqueness does not affect the subsequent analysis.
\end{subequations}
We define the corresponding partition function
\begin{subequations}
\label{eq:bd partition fct nu 4}
\begin{align}
&
Z^{\,}_{\mathrm{bd}}
\df
\int\mathcal{D}[\chi]\int\mathcal{D}[\Phi]\,
e^{
-
S^{\,}_{\mathrm{bd}}
},
\label{eq:bd partition fct nu 4 a}
\\
&
S^{\,}_{\mathrm{bd}}
\df
\int\mathrm{d}^{2}x\,
\left[
\bar{\chi}
\left(
\mathrm{i}\gamma^{\,}_{\mu}
\partial^{\,}_{\mu}
+
\Phi
\right)\chi
+
\frac{1}{(2\lambda)^{2}}
\Phi^{\dag}
\Phi
\right],
\label{eq:bd partition fct nu 4 b}
\end{align}
where, following Ref.\
\onlinecite{PhysRevB.81.045120}, 
we have introduced the notations
\label{eq:Defs nu 4}
\begin{align}
&
\bar{\chi}\df
\chi^{\dag}(-\mathrm{i}\gamma^{\,}_{0}),
\label{eq:bd partition fct nu 4 c}
\\
&
\gamma^{\,}_{0}\df
\beta^{\,}_{4}=\mathrm{X}^{\,}_{102},
\label{eq:bd partition fct nu 4 d}
\\
&
\gamma^{\,}_{1}\df
\mathrm{i}\beta^{\,}_{4}\,\beta^{\,}_{0}=\mathrm{X}^{\,}_{202},
\label{eq:bd partition fct nu 4 e}
\\
&
\gamma^{\,}_{5}\df
\gamma^{\,}_{0}\,\gamma^{\,}_{1}=
\mathrm{i}\beta^{\,}_{0}=
\mathrm{i}\mathrm{X}^{\,}_{300},
\label{eq:bd partition fct nu 4 f}
\\
&
\Upsilon^{\,}_{1}\df
-\phantom{i}\mathrm{X}^{\,}_{312},
\label{eq:bd partition fct nu 4 g}
\\
&
\Upsilon^{\,}_{2}\df
-\phantom{i}\mathrm{X}^{\,}_{332},
\label{eq:bd partition fct nu 4 h}
\\
&
\Upsilon^{\,}_{3}\df
-\phantom{i}\mathrm{X}^{\,}_{320},
\label{eq:bd partition fct nu 4 i}
\\
&
\Upsilon^{\,}_{4}\df
+\mathrm{i}\mathrm{X}^{\,}_{000},
\label{eq:bd partition fct nu 4 j}
\end{align} 
and have defined the matrix-valued field
\begin{align}
&
\Phi(x)\df
|\bm{\phi}(x)|\sum_{l=1}^{4} n^{\,}_{l}(x)\,\Upsilon^{\,}_{l},
\label{eq:bd partition fct nu 4 k}
\\
&
\bm{\phi}(x)\df
|\bm{\phi}(x)|\,\bm{n}(x)
\in\mathbb{R}^{4},
\qquad
\bm{n}^{2}(x)=1,
\label{eq:bd partition fct nu 4 l}
\end{align}
that parametrizes the dynamical 
mass profile. 
We denote  the imaginary time and space coordinates
by $x=(x^{\,}_{0},x^{\,}_{1})
\equiv (\tau,x)$.
\end{subequations}
With the choice of the representation
made in Eqs.\ (\ref{eq:Defs nu 4}),
the identities
\begin{subequations}
\begin{align}
&
\left\{
\gamma^{\,}_{\mu},\gamma^{\,}_{\nu}
\right\}
=
2\delta^{\,}_{\mu\nu},
\qquad
\{\gamma^{\,}_{\mu},\Upsilon^{\,}_{l}\}=
2\delta^{\,}_{4l}\,\Upsilon^{\,}_{l}\,\gamma^{\,}_{\mu},
\label{eq:Commutators nu 4}
\end{align}
\end{subequations}
hold for any $\mu,\nu=0,1$ 
and $l=1,2,3,4$.
Performing the Grassmann integration on
the partition function \eqref{eq:bd partition fct nu 4}
delivers the bosonic and local effective action
\begin{subequations}
\label{eq: functional bosonization for nu=4}
\begin{align}
&
Z^{\,}_{\mathrm{eff}}
\df
\int
\mathcal{D}
\left[
\Phi
\right]
e^{-S^{\,}_{\mathrm{eff}}[\Phi]},
\label{eq: functional bosonization for nu=4 a}
\\
&
S^{\,}_{\mathrm{eff}}[\Phi]
\df
-\frac{1}{2}\,\mathrm{Tr}\,
\left[
\mathrm{ln}\,\mathcal{D}^{\,}_{\Phi}
\right]
+
\frac{1}{32\lambda^{2}}
\mathrm{Tr}\left[\Phi^{\dag}\Phi\right],
\label{eq: functional bosonization for nu=4 b}
\\
&
\mathcal{D}^{\,}_{\Phi}
\df
\mathrm{i}\gamma^{\,}_{\mu}\,\partial^{\,}_{\mu}
+
\Phi.
\label{eq: functional bosonization for nu=4 c}
\end{align}
\end{subequations}
Here, the trace $\mathrm{Tr}$ is understood to be
over both a plane-wave basis and $8\times8$ matrices.
The local effective action 
(\ref{eq: functional bosonization for nu=4 b})
can be written in closed form
to any finite order of a gradient expansion\,\cite{abanov2000theta}
as we now sketch.

The solution $\bar{\Phi}$ to the saddle-point equation
\begin{subequations}
\label{eq:saddle point nu 4}
\begin{align}
\frac{\delta S^{\,}_{\mathrm{eff}}}
{\delta \Phi}=
0
\label{eq:deltaSeff nu 4}
\end{align}
is
\begin{align}
\bar{\Phi}
=
\bar{\phi}
\sum_{\iota=1}^{4}
\bar{n}^{\,}_{\iota}
\Upsilon^{\,}_{\iota},
\end{align}
where
\begin{align}
\sum_{\iota=1}^{4}
\bar{n}^{2}_{\iota}
=1,
\qquad
\bar{\phi}^{2}  
\df
\left(e^{\frac{1}{8\pi\lambda^{2}}}-1\right)^{-1}\,
\Lambda^{2}.
\end{align}
\end{subequations}
Here, $\Lambda$ is the UV cutoff introduced to regularize
the integration over momenta. The direction
of the saddle-point solution $\bar{\bm{n}}$
is arbitrary.

Next, we first consider the change 
$\delta S^{\,}_{\mathrm{eff}}[\Phi]$ of 
effective action 
\eqref{eq: functional bosonization for nu=4}
when $\Phi$ is varied to $\Phi+\delta\Phi$,
\begin{align}
\label{eq:deltaSeff expanded nu 4}
\delta S^{\,}_{\mathrm{eff}}[\Phi]=
S^{\,}_{\mathrm{eff}}[\Phi+\delta\Phi]
-
S^{\,}_{\mathrm{eff}}[\Phi],
\end{align}
which is to be expanded around the saddle-point
solution \eqref{eq:saddle point nu 4}
in powers of $1/\bar{\phi}^{2}$.
Taking the limit $\bar{\phi}^{2}\to \infty$
kills all but a finite number of terms
on the right-hand side of Eq.\ \eqref{eq:deltaSeff expanded nu 4}.
Integration over $\delta \Phi$ then delivers 
two terms.
The first term is
\begin{align}
\label{eq:O(4) nlsm nu 4}
S^{\,}_{\mathrm{NLSM}}
=
\int 
\mathrm{d}^{2}x\,
\frac{1}{2g}
\left(\partial^{\,}_{\mu}\bm{n}\right)^{2},
\qquad
g=\pi.
\end{align}
This is the action of the O(4)-NLSM
in two-dimensional Euclidean spacetime
with the bare coupling constant $g=\pi$.
The second term is
\begin{align}
\Gamma=
\frac{2\mathrm{i}\pi}{3!\mathrm{Area}(\mathsf{S}^{3})}
\int\mathrm{d}^{3}\tilde{x}\,
\epsilon^{\,}_{\mu\nu\rho}\,
\epsilon^{\,}_{abcd}\,
(\partial^{\,}_{\mu}\tilde{n}^{\,}_{a})\,
(\partial^{\,}_{\nu}\tilde{n}^{\,}_{b})\,
(\partial^{\,}_{\rho}\tilde{n}^{\,}_{c})\,
\tilde{n}^{\,}_{d}.
\end{align}
Indeed, imposing
the nonlinear constraint $\bm{n}^{2}(x)=1$
compactifies the target space of the O(4)-NLSM to
the three-sphere $\mathsf{S}^{3}$. This sphere has
the nontrivial homotopy group $\pi^{\,}_{3}(\mathsf{S}^{3})=\mathbb{Z}$.
It is then meaningful following Witten 
\cite{witten1984}
to introduce an auxiliary coordinate
$u\in[0,1]$ and to extend the domain of definition of the field
$\bm{n}(x)$ from $\mathbb{R}^{2}$ to
$\mathbb{R}^{2}\times[0,1]$,
$\bm{n}(x)\to\tilde{\bm{n}}(x,u)$,
$\mathrm{d}^{2}x\to\mathrm{d}^{2}x\,\mathrm{d}u\equiv\mathrm{d}^{3}\tilde{x}$
such that the boundary conditions
$\bar{\bm{n}}(x,0)=\bm{n}^{\,}_{0}$ for some
arbitrary direction $\bm{n}^{\,}_{0}$
and $\tilde{\bm{n}}(x,1)=\bm{n}(x)$ are satisfied. This is the WZ
\cite{WESS197195,Novikov_1982,WITTEN1983422,witten1984}
term for the O(4)-$\mathrm{NLSM}$
in two-dimensional Euclidean space-time.
This term is not local in the action but its effect on the equations of motion
is local. However, this term modifies nonperturbatively
the RG flow obeyed by the coupling $g$.
In fact, in the presence of the WZ term,
the beta function of $g$ has been conjectured to vanish
at the value $g^{\,}_{\mathrm{c}}=\pi$ that defines
a critical point with conformal symmetry\
\cite{witten1984,abanov2000theta}.

The interaction that we chose has an $\mathrm{O}(4)$ symmetry.
This symmetry is not sacred. For example, we
could have introduced four dimensionless couplings
$\lambda^{\,}_{l}$ with $l=1,\cdots,4$,
one for each dynamical mass $\beta^{\,}_{l}$ in
Eqs.\ (\ref{eq:dyn Ham nu 4}).
By treating each dynamical mass $\beta^{\,}_{l}$
as independent Hubbard-Stratonovich fields
and integrating over these fields,
the interaction is the sum of four
quartic contact interactions, each of which is weighted by 
the multiplicative factor $(2\lambda^{\,}_{l})^{-2}$.
This interacting theory can be bosonized with the help of
Abelian bosonization rules. The stability analysis then
proceeds along the same line as what is done in Sec.\ \ref{Sec:haldane}
with the same conclusions. The boundary theory is gapped if and only if
the protecting symmetries
(\ref{eq:imposing DIIR symmetries defs})
or
(\ref{eq:imposing DIIR symmetries defs d})
are spontaneously broken.
One may repeat this exercise with $\nu=6$ and reach the same conclusion,
a gap is necessarily associated with the spontaneous symmetry breaking
of the TRS or RS. It is only when $\nu$ is an integer multiple of the number 8
that a gap delivers a nondegenerate ground state.

\section{The case $\nu=2$}
\label{sec:nu 2}

The set (\ref{eq:ShorthandMatrix defs}) with $n = 2$ has the 16 elements
$\{\mathrm{X}^{\,}_{\mu\rho}\}$ with $\mu,\rho=0,\cdots,3$. 
This set spans the space of $4\times4$ 
Hermitian matrices.
For $\nu=2$, there are at most $\mathrm{N}(2)=2$ interaction channels 
allowed by the symmetry conditions \eqref{eq:imposing DIIR symmetries defs},
each of which is labeled by the Hermitian $4\times4$ matrix $\beta^{\,}_{l}$.
We consider the parametrization
\begin{subequations}
\begin{equation}
\mathcal{H}^{(\mathrm{dyn})}_{\mathrm{bd}}(\tau,x)
\df
\beta^{\,}_{0}\,
\mathrm{i}\partial^{\,}_{x}
+
\sum_{l = 1}^{2}
\beta^{\,}_{l}\,
\phi^{\,}_{l}(\tau,x)
\label{eq:def H bd nu=2}
\end{equation}
of the dynamical boundary single-particle Hamiltonian.
\end{subequations}
Following the same steps as in Sec. \ref{sec:nu 4}, 
we impose the nonlinear constraint:
\begin{align}
\phi^{2}_{1}(\tau,x)+\phi^{2}_{2}(\tau,x)=
\bar{\phi}^{2}.
\end{align}
This condition compactifies the target space of
the effective bosonic and local theory to the circle
$\mathsf{S}^{1}$. However, the nontrivial fundamental group
$\pi^{\,}_{1}(\mathsf{S}^{1})=\mathbb{Z}$ implies the existence of
a topological obstruction.
This topological obstruction takes the form of point defects
when the vector $(\phi^{\,}_{1}(\tau,x),\phi^{\,}_{2}(\tau,x))$ accommodates
vortex configurations in  $(1+1)$-dimensional space-time.
A vortex configuration is singular at the vortex core
where its gradient is ill-defined. 
Direct application of the  gradient expansion method employed in Sec.\ 
\ref{sec:nu 4} is thus invalid.
To circumvent this difficulty, we choose the method of
Abelian bosonization to derive 
an effective local bosonic action.

\subsection{Abelian bosonization}
\label{sec:bosonization}

We start from
\begin{align}
\widehat{H}^{\,}_{\mathrm{bd}}
\df&\,
\int\mathrm{d}x\,
\Big\{
\hat{\chi}^{\dag}
\mathrm{X}^{\,}_{30}\mathrm{i}\partial^{\,}_{x}
\hat{\chi}
-
\sum_{l=1}^{2}
\lambda^{2}_{l}
\left(
\hat{\chi}^{\dag}\,\beta^{\,}_{l}\,\hat{\chi}
\right)^{2}
\Big\},
\label{eq:Hamiltonian interacting nu 2}
\end{align}
i.e., we do not impose the O(2) symmetry
resulting from demanding that
$\lambda^{2}_{1}=\lambda^{2}_{2}=\lambda^{2}$
as is done in Hamiltonian \eqref{eq:Begining Hamiltonian defs}.
Imposing symmetry conditions \eqref{eq:imposing DIIR symmetries defs a}
leads to the identification of 
two possible sets $\left\{\beta^{\,}_{l}\right\}$
\begin{align}
\label{eq:mass sets nu 2}
\mathcal{B}^{\,}_{a}=\{\mathrm{X}^{\,}_{12},\mathrm{X}^{\,}_{20}\},
\quad
\mathcal{B}^{\,}_{b}=\{\mathrm{X}^{\,}_{21},\mathrm{X}^{\,}_{23}\}.
\end{align}
Choosing set
$\mathcal{B}^{\,}_{a}$
in Eq.\ (\ref{eq:Hamiltonian interacting nu 2})
defines $\widehat{H}^{\,}_{\mathrm{bd}\,a}$.
Choosing set
$\mathcal{B}^{\,}_{b}$
in Eq.\ (\ref{eq:Hamiltonian interacting nu 2})
defines $\widehat{H}^{\,}_{\mathrm{bd}\,b}$.
We will perform the subsequent analysis for 
both 
$\widehat{H}^{\,}_{\mathrm{bd}\,a}$
and 
$\widehat{H}^{\,}_{\mathrm{bd}\,b}$
in parallel.
With the convention
\begin{subequations}
\label{eq:Hamiltonian Majorana fermion nu 2}
\begin{equation}
\hat{\chi}^{\dag}=
\left(
\hat{\chi}^{1}_{\mathrm{L}},
\hat{\chi}^{2}_{\mathrm{L}},
\hat{\chi}^{1}_{\mathrm{R}},
\hat{\chi}^{2}_{\mathrm{R}}
\right),
\end{equation}
Hamiltonians
$\widehat{H}^{\,}_{\mathrm{bd}\,a}$
and
$\widehat{H}^{\,}_{\mathrm{bd}\,b}$
are given by
\begin{align}
\widehat{H}^{\,}_{\mathrm{bd}\,a}\df&\,
\int\mathrm{d}x\,
\Big\{
\hat{\chi}^{\dag}
\mathrm{X}^{\,}_{30}\mathrm{i}\partial^{\,}_{x}
\hat{\chi}
\nonumber\\
&\,
-
\lambda^{2}_{1,a}
\left(
\hat{\chi}^{\dag}\mathrm{X}^{\,}_{12}\hat{\chi}
\right)^{2}
-
\lambda^{2}_{2,a}
\left(
\hat{\chi}^{\dag}\mathrm{X}^{\,}_{20}\hat{\chi}
\right)^{2}
\Big\},
\label{eq:Hamiltonian Majorana fermion nu 2 a}
\end{align}
and
\begin{align}
\widehat{H}^{\,}_{\mathrm{bd}\,b}\df&\,
\int\mathrm{d}x\,
\Big\{
\hat{\chi}^{\dag}\,
\mathrm{X}^{\,}_{30}\,
\mathrm{i}\partial^{\,}_{x}\,
\hat{\chi}
\nonumber\\
&\,
-
\lambda^{2}_{1,b}
\left(
\hat{\chi}^{\dag}\,
\mathrm{X}^{\,}_{21}\,
\hat{\chi}
\right)^{2}
-
\lambda^{2}_{2,b}
\left(
\hat{\chi}^{\dag}\,
\mathrm{X}^{\,}_{23}\,
\hat{\chi}
\right)^{2}
\Big\},
\label{eq:Hamiltonian Majorana fermion nu 2 b}
\end{align}
\end{subequations}
respectively. This Majorana representation is not well suited
for Abelian bosonization. 
Instead of it, we define
the right-moving complex fermion fields
\begin{subequations}
\label{eq: changing basis from Majorana to complex}
\begin{equation}
\hat{\psi}^{\dag}_{\mathrm{R}}\df
\frac{
\hat{\chi}^{1}_{\mathrm{R}}
-
\mathrm{i}\hat{\chi}^{2}_{\mathrm{R}}
 }
{\sqrt{2}},
\qquad
\hat{\psi}^{\,}_{\mathrm{R}}\df
\frac{
\hat{\chi}^{1}_{\mathrm{R}}
+
\mathrm{i}\hat{\chi}^{2}_{\mathrm{R}}
 }
{\sqrt{2}},
\label{eq: changing basis from Majorana to complex a}
\end{equation}
the left-moving complex fermion fields
\begin{equation}
\hat{\psi}^{\dag}_{\mathrm{L}}\df
\frac{
\hat{\chi}^{1}_{\mathrm{L}}
-
\mathrm{i}\hat{\chi}^{2}_{\mathrm{L}}
}
{\sqrt{2}},
\qquad
\hat{\psi}^{\,}_{\mathrm{L}}\df
\frac{
\hat{\chi}^{1}_{\mathrm{L}}
+
\mathrm{i}\hat{\chi}^{2}_{\mathrm{L}}
     }
{\sqrt{2}},
\label{eq: changing basis from Majorana to complex b}
\end{equation}
and the complex fermion basis
\begin{equation}
\hat{\Psi}^{\dag}\:=
\begin{pmatrix}
\hat{\psi}^{\dag}_{\mathrm{L}}&
\hat{\psi}^{\dag}_{\mathrm{R}}&
\hat{\psi}^{\,}_{\mathrm{L}}&
\hat{\psi}^{\,}_{\mathrm{R}}
\end{pmatrix}.
\label{eq: changing basis from Majorana to complex c} 
\end{equation}
\end{subequations}
In the basis
(\ref{eq: changing basis from Majorana to complex c}),
we find the complex fermion representation
\begin{subequations}
\label{eq:Hamiltonian complex fermion nu 2}
\begin{align}
\widehat{H}^{\,}_{\mathrm{bd}\,a}\df&\,
\int\mathrm{d}x\,
\Big\{
\hat{\Psi}^{\dag}\,
\mathrm{X}^{\,}_{03}\,
\mathrm{i}\partial^{\,}_{x}\,
\hat{\Psi}
\nonumber\\
&
-
\lambda^{2}_{1,a}
\left(
\hat{\Psi}^{\dag}\,
\mathrm{X}^{\,}_{31}\,
\hat{\Psi}
\right)^{2}
-
\lambda^{2}_{2,a}
\left(
\hat{\Psi}^{\dag}\,
\mathrm{X}^{\,}_{02}\,
\hat{\Psi}
\right)^{2}
\Big\},
\label{eq:Hamiltonian complex fermion nu 2 b}
\end{align}
and
\begin{align}
\widehat{H}^{\,}_{\mathrm{bd}\,b}\df&\,
\int\mathrm{d}x\,
\Big\{
\hat{\Psi}^{\dag}
\mathrm{X}^{\,}_{03}\,
\mathrm{i}\partial^{\,}_{x}\,
\hat{\Psi}
\nonumber\\
&
-
\lambda^{2}_{1,b}
\left(
\hat{\Psi}^{\dag}\,
\mathrm{X}^{\,}_{22}\,
\hat{\Psi}
\right)^{2}
-
\lambda^{2}_{2,b}
\left(
\hat{\Psi}^{\dag}\,
\mathrm{X}^{\,}_{12}\,
\hat{\Psi}
\right)^{2}
\Big\}.
\label{eq:Hamiltonian complex fermion nu 2 c}
\end{align}
\end{subequations}
The change of basis (\ref{eq: changing basis from Majorana to complex})
causes a permutation among the
matrices $\mathrm{X}^{\,}_{\mu\rho}$ with $\mu,\rho=0,1,2,3$.
Hamiltonians \eqref{eq:Hamiltonian complex fermion nu 2 b}
or \eqref{eq:Hamiltonian complex fermion nu 2 c}
are to be normal ordered by using point-splitting
and Wick's theorem. These normal-ordered Hamiltonians
are then bosonized by using the identities
\begin{subequations}
\begin{align}
\label{eq:bosonic field def nu 2}
\hat{\psi}^{\dag}_{\mathrm{R}} (x)
\dfr
\hat{\eta}^{\,}_{\mathrm{R}}\,
\frac{e^{
-\mathrm{i}\hat{\varphi}^{\,}_{\mathrm{R}}(x)}}{\sqrt{2\pi\epsilon}},
\quad
\hat{\psi}^{\dag}_{\mathrm{L}} (x)
\dfr
\hat{\eta}^{\,}_{\mathrm{L}}\,
\frac{e^{
+\mathrm{i}\hat{\varphi}^{\,}_{\mathrm{L}}(x)}}{\sqrt{2\pi\epsilon}},
\end{align}
where $\epsilon$ is a short-distance cutoff.
Hereby, we defined the chiral bosonic fields that
obey the algebra
\begin{align}
\label{eq:chiral bosonic algebra nu 2}
[\hat{\varphi}^{\,}_{\mathrm{R}}(x),
\hat{\varphi}^{\,}_{\mathrm{R}}(x')]
&
=
-
[\hat{\varphi}^{\,}_{\mathrm{L}}(x),
\hat{\varphi}^{\,}_{\mathrm{L}}(x')]
\nonumber\\
&
=
\mathrm{i}\pi\,\mathrm{sgn}(x-x'),
\\
[\hat{\varphi}^{\,}_{\mathrm{R}}(x),\hat{\varphi}^{\,}_{\mathrm{L}}(x')]
&
=
0,
\end{align}
\normalsize
and Klein factors $\hat{\eta}^{\,}_{\mathrm{R/L}}$ 
that obey the algebra
\begin{align}
\left\{
\hat{\eta}^{\,}_{\mathrm{R}},	
\hat{\eta}^{\,}_{\mathrm{R}}
\right\}
=
\left\{
\hat{\eta}^{\,}_{\mathrm{L}},	
\hat{\eta}^{\,}_{\mathrm{L}}
\right\}
=
2,
\quad
\left\{
\hat{\eta}^{\,}_{\mathrm{R}},	
\hat{\eta}^{\,}_{\mathrm{L}}
\right\}
=
0.
\end{align}
\end{subequations}
Hamiltonian \eqref{eq:Hamiltonian complex fermion nu 2 b}
has the bosonic representation
\begin{subequations}
\label{eq:Bos H in chiral fields nu 2}
\begin{align}
\widehat{H}^{\,}_{\mathrm{bd}\,a}
&=
\int\mathrm{d}x\,
\Bigg\{
\frac{1}{2\pi}
\left[
\left(\partial^{\,}_{x}\hat{\varphi}^{\,}_{\mathrm{L}}\right)^{2}
+
\left(\partial^{\,}_{x}\hat{\varphi}^{\,}_{\mathrm{R}}\right)^{2}
\right]
\nonumber\\
&
\qquad
+
\frac{\left(\lambda^{2}_{1,a}+\lambda^{2}_{2,a}\right)}{\pi^{2}}
\left(
\partial^{\,}_{x}\hat{\varphi}^{\,}_{\mathrm{L}}\,
+
\partial^{\,}_{x}\hat{\varphi}^{\,}_{\mathrm{R}}
\right)^{2}
\nonumber\\
&
\qquad
+
\frac{2\left(\lambda^{2}_{1,a}-\lambda^{2}_{2,a}\right)}
{\pi^{2}\epsilon^{2}}
\cos\left(2\,\hat{\varphi}^{\,}_{\mathrm{L}}
+2\,\hat{\varphi}^{\,}_{\mathrm{R}}\right)
\Bigg\}.
\label{eq:Bos H in chiral fields nu 2 a}
\end{align}
Hamiltonian \eqref{eq:Hamiltonian complex fermion nu 2 c}
has the bosonic representation
\begin{align}
\widehat{H}^{\,}_{\mathrm{bd}\,b}
&=
\int\mathrm{d}x\,
\Bigg\{
\frac{1}{2\pi}
\left[
\left(\partial^{\,}_{x}\hat{\varphi}^{\,}_{\mathrm{L}}\right)^{2}
+
\left(\partial^{\,}_{x}\hat{\varphi}^{\,}_{\mathrm{R}}\right)^{2}
\right]
\nonumber\\
&
\qquad
+
\frac{\left(\lambda^{2}_{1,b}+\lambda^{2}_{2,b}\right)}{\pi^{2}}
\left(
\partial^{\,}_{x}\hat{\varphi}^{\,}_{\mathrm{L}}\,
-
\partial^{\,}_{x}\hat{\varphi}^{\,}_{\mathrm{R}}
\right)^{2}
\nonumber\\
&
\qquad
+
\frac{2\left(\lambda^{2}_{1,b}-\lambda^{2}_{2,b}\right)}
{\pi^{2}\epsilon^{2}}
\cos\left(2\,\hat{\varphi}^{\,}_{\mathrm{L}}
-2\,\hat{\varphi}^{\,}_{\mathrm{R}}\right)
\Bigg\}.\label{eq:Bos H in chiral fields nu 2 b}
\end{align}
\end{subequations}
In Hamiltonians \eqref{eq:Bos H in chiral fields nu 2}, we have removed the
Klein factors by diagonalizing the operator
$\mathrm{i}\hat{\eta}^{\,}_{\mathrm{R}}\hat{\eta}^{\,}_{\mathrm{L}}$
and choosing the eigenvalue $+1$ sector in the Klein Hilbert space.
The difference between the two sets
$\mathcal{B}^{\,}_{a}$
and
$\mathcal{B}^{\,}_{b}$
in Eq.\ \eqref{eq:mass sets nu 2}
manifests itself as the sign with which $\hat{\varphi}^{\,}_{\mathrm{R}}$
enters Hamiltonians
\eqref{eq:Bos H in chiral fields nu 2 a}
and
\eqref{eq:Bos H in chiral fields nu 2 b},
respectively.
In Hamiltonian
\eqref{eq:Bos H in chiral fields nu 2 a},
the cosine results from the squares of the
backward-scattering term
$\propto\hat{\psi}^{\dag}_{\mathrm{R}}\hat{\psi}^{\,}_{\mathrm{L}}+\mathrm{H.c.}$
In Hamiltonian
\eqref{eq:Bos H in chiral fields nu 2 b},
the cosine results from the squares of the backward-pairing terms 
$\propto\hat{\psi}^{\dag}_{\mathrm{R}}\hat{\psi}^{\dag}_{\mathrm{L}}
+\mathrm{H.c.}$
In the O(2) symmetric case that is defined by the condition
\begin{equation}
\lambda^{2}_{1,m}=\lambda^{2}_{2,m},
\qquad
m=a,b,
\end{equation}
both cosine interactions vanish and the theory remains gapless.
Away from the O(2) symmetric point, the minima of the cosines
are two-fold degenerate. If the cosines dominate over the
kinetic energy, they open a gap with a two-fold degenerate
manifold of ground states.
Since the dependence on
interaction strengths have the same form
in  Hamiltonians
\eqref{eq:Bos H in chiral fields nu 2 a}
and
\eqref{eq:Bos H in chiral fields nu 2 b},
the boundaries in the corresponding phase diagrams
are identical.
However, the phases they separate can be different
whenever they break spontaneously distinct symmetries.

The transformation
\footnote{
 Recall that two copies of the helical Majorana fields can be thought
of as a low energy description of two copies of the Ising model.
Suppose now that a Kramers-Wannier duality transformation 
is applied to only the second copy of the Ising model
via the transformation 
$\hat{\chi}^{2}_{\mathrm{L}}\to\hat{\chi}^{2}_{\mathrm{L}}$ and 
$\hat{\chi}^{2}_{\mathrm{R}}\to-\hat{\chi}^{2}_{\mathrm{R}}.$
In the language	of complex fermions, the 
left-handed component $\hat{\psi}^{\,}_{\mathrm{L}}$ is unchanged, 
while the right-handed component $\hat{\psi}^{\,}_{\mathrm{R}}$ 
is transformed into its dagger, i.e.,
$\hat{\psi}^{\,}_{\mathrm{R}}\to\hat{\psi}^{\dag}_{\mathrm{R}}$.
The transformation (\ref{eq:def KW transformation nu 2}) 
of chiral bosons then follows.
         }
\begin{equation}
\hat{\varphi}^{\,}_{\mathrm{L}}\to+\hat{\varphi}^{\,}_{\mathrm{L}},
\qquad
\hat{\varphi}^{\,}_{\mathrm{R}}\to-\hat{\varphi}^{\,}_{\mathrm{R}}
\label{eq:def KW transformation nu 2}
\end{equation}
that interchanges Hamiltonians
\eqref{eq:Bos H in chiral fields nu 2 a}
and
\eqref{eq:Bos H in chiral fields nu 2 b}
is nothing but the transformation 
that interchanges the pair of dual fields
\begin{subequations}
\begin{align}
\label{eq:def dual fields nu 2}
&\hat{\phi}(x)
\df
\frac{1}{\sqrt{4\pi}}
\left[
\hat{\varphi}^{\,}_{\mathrm{L}}(x)
+
\hat{\varphi}^{\,}_{\mathrm{R}}(x)
\right],
\\
&\hat{\theta}(x)
\df
\frac{1}{\sqrt{4\pi}}
\left[
\hat{\varphi}^{\,}_{\mathrm{L}}(x)
-
\hat{\varphi}^{\,}_{\mathrm{R}}(x)
\right],
\end{align}
that satisfy the algebra
\begin{equation}
\left[
\hat{\phi}(x),
\hat{\theta}(x')
\right]=
\frac{\mathrm{i}}{2}\mathrm{sgn}(x'-x)
\end{equation}
\end{subequations} 
with  all other commutators vanishing. 
If one trades the Hamiltonian representation for
the Lagrangian representation,
one obtains the pair of actions
\begin{subequations}
\label{eq:Bosonic actions nu 2}
\begin{align}
\label{eq:action a nu 2}
S^{\,}_{a} 
&\df
\int \mathrm{d}^{2}x\,
\left\{
\frac{1}{2g^{\,}_{a}}
\left(\partial^{\,}_{\mu}\phi\right)^{2}
+
\kappa^{\,}_{a}
\mathrm{cos}
\left(\sqrt{16\pi}\phi\right)
\right\},
\\
\label{eq:action b nu 2}
S^{\,}_{b} 
&\df
\int \mathrm{d}^{2}x\,
\left\{
\frac{1}{2g^{\,}_{b}}
\left(\partial^{\,}_{\mu}\theta\,\right)^{2}
+
\kappa^{\,}_{b}
\mathrm{cos}
\left(\sqrt{16\pi}\theta\,\right)
\right\},
\end{align}
where $\phi$ and $\theta$ are dual scalar fields satisfying
either
\begin{align}
\partial^{\,}_{\mu}\phi
=
\mathrm{i}\,
g^{\,}_{a}\,
\epsilon^{\,}_{\mu\nu}
\partial^{\,}_{\nu}\theta,
\end{align}
with $\mu=0,1$, $(x^{\,}_{0},x^{\,}_{1})=(v^{\,}_{a}\tau,x)$,
or
\begin{align}
\partial^{\,}_{\mu}\phi
=
\mathrm{i}\,
g^{\,}_{b}\,
\epsilon^{\,}_{\mu\nu}
\partial^{\,}_{\nu}\theta,
\end{align}
with $\mu=0,1$, $(x^{\,}_{0},x^{\,}_{1})=(v^{\,}_{b}\tau,x)$,
respectively.
The coupling constants are given by 
\begin{align}
\label{eq:Coupling constants nu 2 a}
&
\frac{2}{v^{\,}_{a}}
=
g^{\,}_{a} \df 
\frac{1}{
\sqrt{1+
4
\frac{\lambda^{2}_{1,a}+\lambda^{2}_{2,a}}{\pi}}},
\\
\label{eq:Coupling constants nu 2 b}
&
\frac{2}{v^{\,}_{b}}
=
g^{\,}_{b} 
\df 
\frac{1}{
\sqrt{1+
4
\frac{\lambda^{2}_{1,b}+\lambda^{2}_{2,b}}{\pi}}},
\end{align}
whereas the effective interaction strengths are
\begin{align}
\label{eq:Interaction strengths nu 2 a}
&
\kappa^{\,}_{a}
\df
\frac{4}{\pi^{2}\epsilon^{2}}
\sqrt{
1
+
4
\frac{\lambda^{2}_{1,a}+\lambda^{2}_{2,a}}{\pi}}
\left(\lambda^{2}_{1,a}-\lambda^{2}_{2,a}\right),
\\
\label{eq:Interaction strengths nu 2 b}
&
\kappa^{\,}_{b}
\df
\frac{4}{\pi^{2}\epsilon^{2}}
\sqrt{
1
+
4
\frac{\lambda^{2}_{1,b}+\lambda^{2}_{2,b}}{\pi}}
\,\left(\lambda^{2}_{1,b}-\lambda^{2}_{2,b}\,\right).
\end{align}
\end{subequations}
The two actions
\eqref{eq:action a nu 2}
and
\eqref{eq:action b nu 2}
are exchanged if one performs the interchanges
$\lambda^{2}_{i,a}\leftrightarrow\lambda^{2}_{i,b}$
with $i=1,2$  
and
$\phi\leftrightarrow\theta$.
The interaction strengths
\eqref{eq:Interaction strengths nu 2 a}
and \eqref{eq:Interaction strengths nu 2 b}
change signs depending on whether $\lambda^{2}_{1,m}
>\lambda^{2}_{2,m}$ or $\lambda^{2}_{1,m}
<\lambda^{2}_{2,m}$, with $m=a,b$.

Before proceeding, we determine how the symmetries defined in
Eqs.\ \eqref{eq:SymmetryOperators defs} act
on the bosonic fields. The actions of the symmetry transformations
on the complex fermionic fields are deduced
from their actions on the Majorana fields and
given by
\begin{subequations}
\begin{align}
\label{eq:PHSonFields nu 2}
&
\widehat{U}^{\dag}_{\mathrm{C}}\,
\begin{pmatrix}
\,\hat{\psi}^{\,}_{\mathrm{L}}(\tau,x)\\
\,\hat{\psi}^{\,}_{\mathrm{R}}(\tau,x)
\,\end{pmatrix}
\widehat{U}^{\,}_{\mathrm{C}}
=
\begin{pmatrix}
\,\hat{\psi}^{\,}_{\mathrm{L}}(\tau,x)\\
\,\hat{\psi}^{\,}_{\mathrm{R}}(\tau,x)
\end{pmatrix},
\\
\label{eq:TRSonFields nu 2}
&
\widehat{U}^{\dag}_{\mathrm{T}}\,
\begin{pmatrix}
\,\hat{\psi}^{\,}_{\mathrm{L}}(\tau,x)\\
\,\hat{\psi}^{\,}_{\mathrm{R}}(\tau,x)
\,\end{pmatrix}
\widehat{U}^{\,}_{\mathrm{T}}
=
\begin{pmatrix}
\,+\hat{\psi}^{\dag}_{\mathrm{R}}(\tau,x)\\
\,-\hat{\psi}^{\dag}_{\mathrm{L}}(\tau,x)
\end{pmatrix},
\\
\label{eq:RSonFields nu 2}
&
\widehat{U}^{\dag}_{\mathrm{R}}\,
\begin{pmatrix}
\,\hat{\psi}^{\,}_{\mathrm{L}}(\tau,x)\\
\,\hat{\psi}^{\,}_{\mathrm{R}}(\tau,x)
\,\end{pmatrix}
\widehat{U}^{\,}_{\mathrm{R}}
=
\begin{pmatrix}
\,+\hat{\psi}^{\,}_{\mathrm{R}}(\tau,-x)\\
\,-\hat{\psi}^{\,}_{\mathrm{L}}(\tau,-x)
\end{pmatrix},
\end{align}
\end{subequations}
where $\widehat{U}^{\,}_{\mathrm{C}}$, 
$\widehat{U}^{\,}_{\mathrm{T}}$, and $\widehat{U}^{\,}_{\mathrm{R}}$
are PH, reversal of time, and reflection 
transformations at the many-body level.
The operator $\widehat{U}^{\,}_{\mathrm{T}}$ is 
defined to be antiunitary,
whereas operators $\widehat{U}^{\,}_{\mathrm{C}}$
and $\widehat{U}^{\,}_{\mathrm{R}}$
are chosen to be unitary.
We note that the PHS is
represented by the identity, whereas the 
TRS involves a PH transformation
\footnote{
In fact, the unitary particle-hole transformation operator
\protect{$\widehat{U}^{\,}_{\mathrm{C}}$}
replaces the spinor $\hat{\Psi}$ with its conjugate transpose
$\hat{\Psi}^{\dag}(\tau,x)$ by the transformation rule
\protect{
$\widehat{U}^{\dag}_{\mathrm{C}}\,
\hat{\Psi}(\tau,x)\,
\widehat{U}^{\,}_{\mathrm{C}}=
\hat{\Psi}^{\dag}(\tau,x)\,
\mathrm{M}$},
where $\mathrm{M}$ is a unitary matrix.
It follows from the Majorana reality condition 
\eqref{eq:majorana condition defs} and 
the representation \eqref{eq:imposing DIIR symmetries defs a}
that $\mathrm{M}=\mathrm{X}^{\,}_{10}$. This implies
the transformation rule \eqref{eq:PHSonFields nu 2}
for the individual components of the spinor $\hat{\Psi}(\tau,x)$.}.
These transformation laws
together with Eqs.\ (\ref{eq:bosonic field def nu 2})
imply the transformation laws
\begin{subequations}
\label{eq:Sym chiral fields nu 2}
\begin{align}
\label{eq:Sym chiral fields nu 2 a}
&
\widehat{U}^{\dag}_{\mathrm{C}}\,
\begin{pmatrix}
\,\hat{\varphi}^{\,}_{\mathrm{L}}(\tau,x)
\\
\,\hat{\varphi}^{\,}_{\mathrm{R}}(\tau,x)
\,\end{pmatrix}
\widehat{U}^{\,}_{\mathrm{C}}
=
\begin{pmatrix}
\,\hat{\varphi}^{\,}_{\mathrm{L}}(\tau,x)
\\
\,\hat{\varphi}^{\,}_{\mathrm{R}}(\tau,x)
\,\end{pmatrix},
\\
\label{eq:Sym chiral fields nu 2 b}
&
\widehat{U}^{\dag}_{\mathrm{T}}\,
\begin{pmatrix}
\,\hat{\varphi}^{\,}_{\mathrm{L}}(\tau,x)
\\
\,\hat{\varphi}^{\,}_{\mathrm{R}}(\tau,x)
\,\end{pmatrix}
\widehat{U}^{\,}_{\mathrm{T}}
=
\begin{pmatrix}
-\hat{\varphi}^{\,}_{\mathrm{R}}(\tau,x)
\\
-\hat{\varphi}^{\,}_{\mathrm{L}}(\tau,x)+\pi
\,\end{pmatrix},
\\
\label{eq:Sym chiral fields nu 2 c}
&
\widehat{U}^{\dag}_{\mathrm{R}}\,
\begin{pmatrix}
\,\hat{\varphi}^{\,}_{\mathrm{L}}(\tau,x)
\\
\,\hat{\varphi}^{\,}_{\mathrm{R}}(\tau,x)
\,\end{pmatrix}
\widehat{U}^{\,}_{\mathrm{R}}
=
\begin{pmatrix}
-\hat{\varphi}^{\,}_{\mathrm{R}}(\tau,-x)
\\
-\hat{\varphi}^{\,}_{\mathrm{L}}(\tau,-x)
\,\end{pmatrix}.
\end{align}
\end{subequations}
We note that in deriving transformation rules (\ref{eq:Sym chiral fields nu 2}),
one must take care of the transformation rules on the Klein factors as well.
Demanding the invariance of the operator 
$\mathrm{i}\hat{\eta}^{\,}_{\mathrm{R}}\hat{\eta}^{\,}_{\mathrm{L}}$,
we find the transformation rules
\begin{subequations}
\begin{align}
&
\widehat{U}^{\dag}_{\mathrm{T}}\,
\begin{pmatrix}
\,\hat{\eta}^{\,}_{\mathrm{L}}
\\
\,\hat{\eta}^{\,}_{\mathrm{R}}
\,\end{pmatrix}
\widehat{U}^{\,}_{\mathrm{T}}
=
\begin{pmatrix}
\,+\hat{\eta}^{\,}_{\mathrm{R}}
\\
\,+\hat{\eta}^{\,}_{\mathrm{L}}
\end{pmatrix},
\\
&
\widehat{U}^{\dag}_{\mathrm{R}}\,
\begin{pmatrix}
\,\hat{\eta}^{\,}_{\mathrm{L}}
\\
\,\hat{\eta}^{\,}_{\mathrm{R}}
\,\end{pmatrix}
\widehat{U}^{\,}_{\mathrm{R}}
=
\begin{pmatrix}
\,+\hat{\eta}^{\,}_{\mathrm{R}}
\\
\,-\hat{\eta}^{\,}_{\mathrm{L}}
\end{pmatrix}.
\end{align}
\end{subequations}
The corresponding transformation rules for the
bosonic pair of dual fields are then
found to be
\begin{subequations}
\label{eq:Sym on BFields nu 2}
\begin{align}
\label{eq:PHSonBFields nu 2}
&
\widehat{U}^{\dag}_{\mathrm{C}}\,
\begin{pmatrix}
\,\hat{\phi}(\tau,x)
\\
\,\hat{\theta}(\tau,x)
\,\end{pmatrix}
\widehat{U}^{\,}_{\mathrm{C}}
=
\begin{pmatrix}
\,\hat{\phi}(\tau,x)
\\
\,\hat{\theta}(\tau,x)
\end{pmatrix},
\\
\label{eq:TRSonBFields nu 2}
&
\widehat{U}^{\dag}_{\mathrm{T}}\,
\begin{pmatrix}
\,\hat{\phi}(\tau,x)
\\
\,\hat{\theta}(\tau,x)
\,\end{pmatrix}
\widehat{U}^{\,}_{\mathrm{T}}
=
\begin{pmatrix}
-\hat{\phi}(\tau,x)+\sqrt{\pi}/2
\\
\,+\hat{\theta}(\tau,x)-\sqrt{\pi}/2
\end{pmatrix},
\\
\label{eq:RSonBFields nu 2}
&
\widehat{U}^{\dag}_{\mathrm{R}}\,
\begin{pmatrix}
\,\hat{\phi}(\tau,x)
\\
\,\hat{\theta}(\tau,x)
\,\end{pmatrix}
\widehat{U}^{\,}_{\mathrm{R}}
=
\begin{pmatrix}
\,-\hat{\phi}(\tau,-x)
\\
\,+\hat{\theta}(\tau,-x)
\end{pmatrix}.
\end{align}
\end{subequations}
Alternatively, the transformations (\ref{eq:Sym on BFields nu 2})
can also be deduced from applying the many-body symmetry transformations on
the components of the fermionic two-current.

Equipped with the transformation rules (\ref{eq:Sym on BFields nu 2}), 
we explore the phase diagram corresponding
to the actions \eqref{eq:Bosonic actions nu 2}.
For both actions, the corresponding cosine term has 
the scaling dimension
\begin{align}
\Delta^{\,}_{m}
\df
\frac{4}{\sqrt{1
+
4
\frac{\lambda^{2}_{1,m}+\lambda^{2}_{2,m}}{\pi}}},
\quad
m=a,b.
\end{align}
Therefore, the cosine terms are IR irrelevant 
when $\lambda^{2}_{1,m}+\lambda^{2}_{2,m}<3\pi/4$ 
and the theory remains critical.
Increasing the interaction strengths makes the cosines relevant, 
in which case the fields $\theta$ and $\phi$ 
are pinned to the minima of the corresponding
cosine terms in the ground state.

Each cosine has four extrema, two of which
become minima depending on the difference 
$\lambda^{2}_{1,m}-\lambda^{2}_{2,m}$ being positive 
or negative. In particular, when this difference is 
zero, both cosines vanish and the low-energy effective
theory is that of a free scalar field, i.e.,
it also remains critical. 
This is the O(2)-symmetric line in the parameter space. 
Away from this line, 
we observe  twofold ground-state degeneracy due to 
the two minima of the cosine. 

For action \eqref{eq:action a nu 2}
with $\lambda^{2}_{1,a}>\lambda^{2}_{2,a}$,
the two ground states are $\phi=\sqrt{\pi}/4$ and
$\phi=3\sqrt{\pi}/4$.
The transformation
rules \eqref{eq:Sym on BFields nu 2} then imply that 
RS is spontaneously broken. 
Conversely, when $\lambda^{2}_{1,a}<\lambda^{2}_{2,a}$,
the ground states correspond to $\phi=0$
and  $\phi=\sqrt{\pi}/2$,
which implies that TRS is spontaneously broken.

For action \eqref{eq:action b nu 2},
the transformation rules \eqref{eq:Sym on BFields nu 2} 
imply that RS always holds,
whereas TRS is broken whenever
there are two ground states separated by a shift of
$\hat{\theta}$ by $\sqrt{\pi}/2$. This is realized by
the cosine interaction in Eq.\ 
\eqref{eq:action b nu 2}. 

\begin{figure}[t]
\centering
\includegraphics[width=0.49\textwidth]{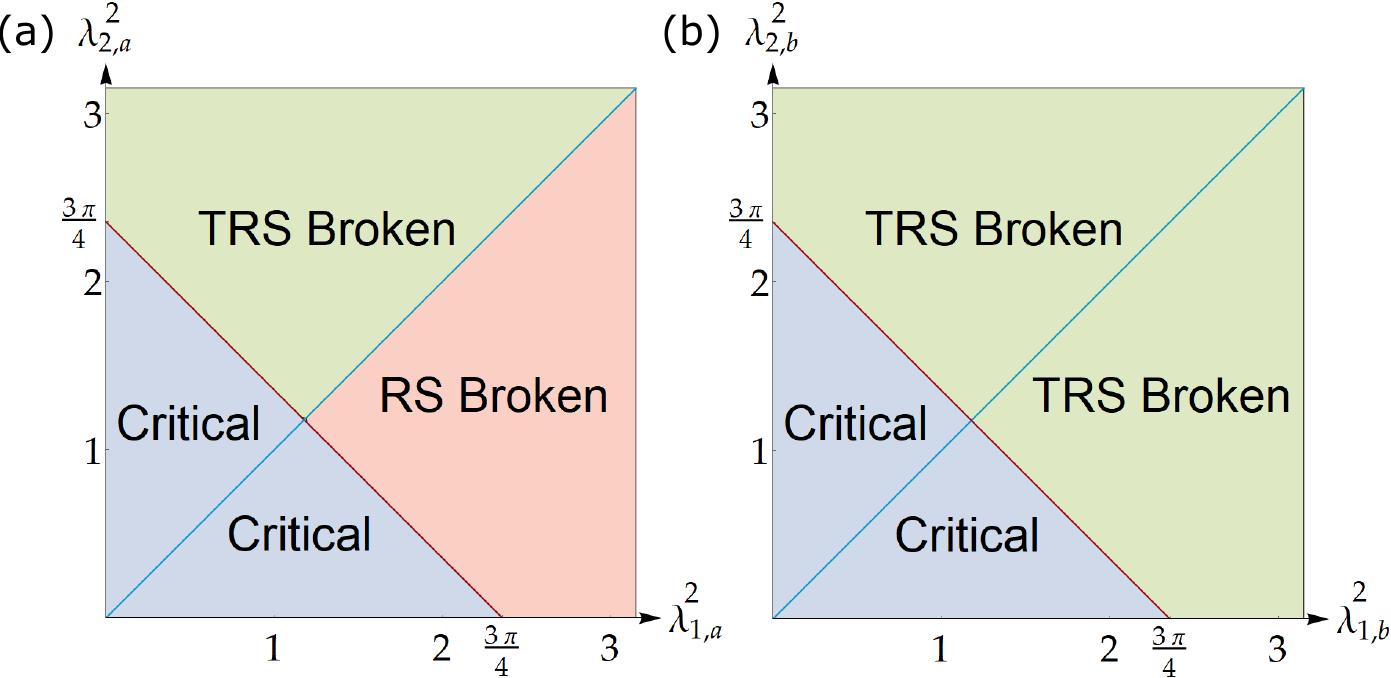}
\caption{
Phase diagram for the edge theories defined
by the actions
\eqref{eq:action a nu 2} ($m=a$) in panel (a)
and
\eqref{eq:action b nu 2} ($m=b$) in panel (b)
as a function of the interaction strengths $\lambda^{2}_{i,m}$ with
$i=1,2$ and $m=a,b$.
Along the blue line,
O(2) symmetry holds and both cosine interactions vanish.
Along the red line, both cosine interactions are marginal.
}
\label{fig:phase diagram nu 2}
\end{figure}

In Fig.\ \ref{fig:phase diagram nu 2},
we plot the phase diagrams for both actions
\eqref{eq:action a nu 2} ($m=a$)
and
\eqref{eq:action b nu 2} ($m=b$)
as functions of the interaction strengths
$\lambda^{2}_{1,m}$ and $\lambda^{2}_{2,m}$,
respectively.
For given $m=a,b$, we define the red line
in Fig.\ \ref{fig:phase diagram nu 2} by 
\begin{equation}
\lambda^{2}_{1,m}+\lambda^{2}_{2,m}=3\pi/4
\ \Longleftrightarrow\
\Delta^{\,}_{m}=2
\label{eq: def red line}
\end{equation}
and the blue line in Fig.\ \ref{fig:phase diagram nu 2} by 
\begin{equation}
\lambda^{2}_{1,m}=\lambda^{2}_{2,m}.
\label{eq: def blue line}
\end{equation}
Below the red line (\ref{eq: def red line}),
the cosine interactions are irrelevant
as their scaling dimensions are larger than 2.
Each point in coupling space is then a critical phase
with algebraic correlation functions
characterized by scaling exponents that are smooth functions
of the couplings $\lambda^{2}_{1,m}$ and $\lambda^{2}_{2,m}$.
The free Dirac point is defined by the origin
$\lambda^{2}_{1,m}=\lambda^{2}_{2,m}=0$
of coupling space.
Above the red line (\ref{eq: def red line}),
the cosine interactions are relevant
as their scaling dimensions are smaller than 2. 
Each point in coupling space then belongs to a gapped phase,
unless the couplings multiplying the cosine interactions vanish,
as they do along the blue line (\ref{eq: def blue line}).
Each gapped phase is associated with a pattern of spontaneous
symmetry breaking. When $\lambda^{2}_{1,a}<\lambda^{2}_{2,a}$ 
($\lambda^{2}_{1,a}>\lambda^{2}_{2,a}$), TRS (RS) is spontaneously
broken as follows from minimizing the cosine interaction.
When $\lambda^{2}_{1,b}\neq\lambda^{2}_{2,b}$, TRS  is spontaneously
broken as follows again from minimizing the cosine interaction.
Along the blue line (\ref{eq: def blue line}),
O(2) symmetry holds and both cosine interactions vanish.
Along the red line, both cosine interactions are marginal.

Abelian bosonization reveals that when quartic contact
interactions compatible with the DIIIR symmetries are 
added, gap opening necessarily breaks one of the 
defining symmetries. Therefore, the $\nu=2$ edge theory
remains stable in the presence of interactions
in the sense that it may only be gapped by interactions if
any one of the protecting symmetries is either explicitly
or spontaneously broken.
We will next consider a generic family
of symmetry-preserving cosine interactions and demonstrate that
any interaction that gaps the edge theory must necessarily
break spontaneously one of the protecting symmetries.
We also discuss the effect of breaking of TS. 

\subsection{Haldane criterion}
\label{Sec:haldane}

In Sec. \ref{sec:bosonization},
we bosonized Hamiltonians 
\eqref{eq:Hamiltonian complex fermion nu 2}.
They are of the sine-Gordon type.
In light of this result, one may 
consider a family of bosonic Hamiltonians
with generic cosine interactions.
These interactions can gap some, most, or all bosonic degrees of freedom.
How many bosonic degrees of freedom remain gapless
is determined using the so-called
Haldane stability criterion~{\cite{PhysRevLett.74.2090}}.
Doing so in a manner compliant with imposing the protecting symmetries,
we are going to recover the cosine potentials 
\eqref{eq:Bos H in chiral fields nu 2}.

We consider the 
Hamiltonian
\begin{subequations}
\label{eq:Chiral bosonic theory haldane}
\begin{align}
\widehat{H}
\df
\widehat{H}^{\,}_{0}
+
\widehat{H}^{\,}_{\mathrm{int}},
\label{eq:Chiral bosonic theory haldane a}
\end{align}
which consists of the free Hamiltonian
\begin{align}
\widehat{H}^{\,}_{0}
\df
\int
\mathrm{d}x\,
\frac{1}{4\pi}
\left(
\partial^{\,}_{x}
\widehat\Phi^{\mathsf{T}}
\right)
(x)
\,V\,
\left(
\partial^{\,}_{x}
\widehat\Phi
\right)
(x),
\label{eq:chiral bosonic free haldane}
\end{align}
that describes free chiral bosonic fields 
and the interaction
\begin{align}
\widehat{H}^{\,}_{\mathrm{int}}\df
- 
\int\!
\mathrm{d}x\!
\sum_{T\in\mathbb{H}}
h^{\,}_{T}(x)
\bm{:}\!
\cos\left(T^{\mathsf{T}}\,K\,
\widehat\Phi(x)+\alpha^{\,}_{T}(x)\right)\!
\bm{:}
\label{eq:chiral bosonic int haldane}
\end{align}
that encodes
a countable set of local fermionic interactions
describing many-body umklapp processes
that we shall call tunneling processes and hence label with the symbol $T$.  
The components of the field $\widehat\Phi$ obey
the commutation relations
\begin{align}
\left[
\widehat\Phi^{\,}_{i}(x),
\widehat\Phi^{\,}_{j}(x')
\right]
=
-\mathrm{i}\pi
\left[
K^{-1}_{ij}
\mathrm{sgn}(x-x')
\right],
\label{eq:chiral bosonic algebra haldane}
\end{align}
where $K$ is a $2\times2$, integer valued, symmetric, and
invertible matrix. 
The static functions
\begin{equation}
h^{\,}_{T}(x)\geq 0,
\qquad
0\leq\alpha^{\,}_{T}(x)<2\pi
\end{equation}
\end{subequations}
encode the possibility that TS is broken on the edge.
The matrix $V$ is a $2\times2$ symmetric and positive definite matrix.
The two-dimensional tunneling vectors $T$ 
are chosen from a set $\mathbb{H}$, that we
will specify later.

Our aim is to compare Hamiltonian
\eqref{eq:Chiral bosonic theory haldane a}
with Hamiltonian \eqref{eq:Bos H in chiral fields nu 2 a}
or Hamiltonian \eqref{eq:Bos H in chiral fields nu 2 b}
and use the Haldane criterion to identify some ``minimal''
sets of tunneling vectors $\mathbb{H}$ 
that would gap the chiral bosonic fields $\widehat\Phi$
if the functions $h^{\,}_{\mathrm{T}}$ were ``large''.
By comparing the free Hamiltonian
\eqref{eq:chiral bosonic free haldane}
with \eqref{eq:Bos H in chiral fields nu 2},
we define the fields
\begin{subequations}
\label{eq:input data haldane}
\begin{align}
\widehat\Phi(x)
\df
\begin{pmatrix}
\hat{\varphi}^{\,}_{\mathrm{L}}
\left(x\right)&
\hat{\varphi}^{\,}_{\mathrm{R}}
\left(x\right)
\end{pmatrix}^{\mathsf{T}},
\label{eq:input data haldane a}
\end{align}
the universal data
\begin{align}
Q\df 
\begin{pmatrix}
1
\\
1
\end{pmatrix},
\quad
K\df
\begin{pmatrix}
+1 & \phantom{-}0
\\
\phantom{+}0 & -1
\end{pmatrix},
\label{eq:input data haldane b}
\end{align}
and the nonuniversal data
\begin{equation}
V 
\df
\begin{pmatrix}
v & u\\
u & v
\end{pmatrix},
\qquad
0<v\in\mathbb{R},
\qquad
0\leq u\in\mathbb{R}.
\label{eq:input data haldane c}
\end{equation}
\end{subequations}
With the universal  data \eqref{eq:input data haldane b},
the algebra \eqref{eq:chiral bosonic algebra haldane}
reduces to the algebra \eqref{eq:chiral bosonic algebra nu 2}.
The two-dimensional vector $Q$ is
the charge vector. The explicit dependence
of the positive couplings $u$ and $v$ on the couplings
$\lambda^{2}_{i,m}$, $i=1,2$, $m=a,b$
from Hamiltonian \eqref{eq:Bos H in chiral fields nu 2 a}
will not be needed in the following.

The minimal set of tunneling vectors $\mathbb{H}$
is defined as follows. 
We first construct the maximal Haldane set
\begin{align}
\mathbb{L}\df
\left\{
T\in \mathbb{Z}^{2}
\,\left|\,
T^{\mathsf{T}}\,
K\,
T^{\prime}
=
T^{\prime\mathsf{T}}\,
K\,
T
=
0,
\forall T'\in \mathbb{L}
\right.\right\},
\label{eq:def tunneling vectors set haldane}
\end{align}
i.e.,
the set of elements in $\mathbb{Z}^{2}=\mathbb{Z}\times\mathbb{Z}$
such that the bilinear form
$T^{\mathsf{T}}\,K\,T^{\prime}$ vanishes
for any pair $T$ and $T^{\prime}$ from $\mathbb{L}$.
This constraint is the compatibility condition of the Haldane criterion.
With it, there is no competition between any pair of
cosine interaction entering $\widehat{H}^{\,}_{\mathrm{int}}$.
The vectors $T\in \mathbb{L}$  form a lattice since,
for any pair $T,T'\in\mathbb{L}$, the linear combination
$n\,T+n'\,T'$ with $n,n'\in\mathbb{Z}$
also satisfies the compatibility condition.
We then define the minimal set of tunneling vectors
as the subset $\mathbb{H}\subset\mathbb{L}$
such that elements $T\in\mathbb{H}$
constitutes the primitive cell of the lattice 
$\mathbb{L}$ which is compatible
with the symmetry requirements of class DIIIR.

The
Haldane criterion then asserts that the Hamiltonian
\eqref{eq:chiral bosonic int haldane} for a given $\mathbb{H}$,
removes $2\times|\mathbb{H}|$-chiral 
bosonic fields from gapless degrees of freedom by pinning them
(the notation $|\mathbb{H}|$
denotes the cardinality of the set $\mathbb{H}$).
In our case, Hamiltonian \eqref{eq:Chiral bosonic theory haldane a}
consists of only a single pair of chiral bosonic fields.
Therefore, it is enough to find the single tunneling vector
making up $\mathbb{H}$ to remove all gapless degrees of freedom. 

For a general tunneling vector
$T = (m,n)$ of integers $m,n\in\mathbb{Z}$,
the Haldane compatibility condition implies
that there are two solutions, $n=m$
and $n=-m$. Therefore, there exists two
disjoint sets of lattices $\mathbb{L}$
generated by the primitive cells
\begin{subequations}
\label{eq:maximal tunneling set haldane}
\begin{align}
\label{eq:maximal tunneling set haldane a}
&
\mathbb{H}^{\,}_{a}\df
\left\{\left.
\begin{pmatrix}
+n^{\,}_{a}&-n^{\,}_{a}\end{pmatrix}^{\mathsf{T}}\,\right|\,
n^{\,}_{a} \hbox{ to be determined}
\right\},
\\
\label{eq:maximal tunneling set haldane b}
&
\mathbb{H}^{\,}_{b}\df
\left\{\left.
\begin{pmatrix}+n^{\,}_{b}&+n^{\,}_{b}\end{pmatrix}^{\mathsf{T}}\,\right|\,
n^{\,}_{b} \hbox{ to be determined}
\right\}.
\end{align}
\end{subequations}
The integers $n^{\,}_{a}$ and $n^{\,}_{b}$ are not yet determined.
To determine how integers $n^{\,}_{a}$
and $n^{\,}_{b}$ are constrained, we define the pair of 
interactions
\begin{subequations}
\label{eq:H int haldane}
\begin{align}
\label{eq:H int haldane a}
\widehat{H}^{\,}_{\mathrm{int}\,a}\df
\int\!
\mathrm{d}x\,
h^{\,}_{a}(x)
\bm{:}\!
\cos\big(
n^{\,}_{a}\left[
\hat{\varphi}^{\,}_{\mathrm{L}}
+\hat{\varphi}^{\,}_{\mathrm{R}}
\right](x)
+
\alpha^{\,}_{a}(x)\big)\!
\bm{:},
\end{align}
and
\begin{align}
\label{eq:H int haldane b}
\widehat{H}^{\,}_{\mathrm{int}\,b}\df
\int\!
\mathrm{d}x\,
h^{\,}_{b}(x)
\bm{:}\!
\cos\big(
n^{\,}_{b}
\left[
\hat{\varphi}^{\,}_{\mathrm{L}}
-
\hat{\varphi}^{\,}_{\mathrm{R}}
\right](x)
+
\alpha^{\,}_{b}(x)\big)\!
\bm{:},
\end{align}
\end{subequations}
corresponding to the minimal sets 
\eqref{eq:maximal tunneling set haldane a}
and \eqref{eq:maximal tunneling set haldane b}, respectively,
on which we shall impose the 
symmetries under the transformations
defined in Eq.\
\eqref{eq:Sym on BFields nu 2}.
Observe that, in the strong coupling limit
\begin{subequations}
\label{eq: H int minima}
\begin{equation}
4\pi\,\mathrm{sup}\{h^{\,}_{a}(x)\}\gg\max\{u,v\}
\end{equation}
[recall that $h^{\,}_{a}(x)\geq0$ for any $x$
and $u$ and $v$ are defined in the velocity matrix
(\ref{eq:input data haldane c})], 
the linear combinations
$\hat{\varphi}^{\,}_{\mathrm{L}}(x)\pm\hat{\varphi}^{\,}_{\mathrm{R}}(x)$
of the chiral fields are pinned to the minima of the cosine potentials, namely,
either
\begin{align}
n^{\,}_{a}
\left[
\hat{\varphi}^{\,}_{\mathrm{L}}(x)
+
\hat{\varphi}^{\,}_{\mathrm{R}}(x)
\right]
=
2\pi k 
+
\pi 
-
\alpha^{\,}_{a}(x),
\label{eq: H int minima a}
\end{align}
or
\begin{align}
n^{\,}_{b}
\left[
\hat{\varphi}^{\,}_{\mathrm{L}}(x)
-
\hat{\varphi}^{\,}_{\mathrm{R}}(x)
\right]
=
2\pi k 
+
\pi 
-
\alpha^{\,}_{b}(x),
\label{eq: H int minima b}
\end{align}
\end{subequations}
respectively, for some integer $k\in\mathbb{Z}$.

\subsubsection{Symmetry constraints on Hamiltonian \eqref{eq:H int haldane a}}

PHS is trivially satisfied by construction.
Imposing TRS by using the transformation rule
\eqref{eq:Sym chiral fields nu 2 b}
leads to the constraint
\begin{subequations}
\begin{equation}
\alpha^{\,}_{a}(x)
=
-
\alpha^{\,}_{a}(x)
-
n^{\,}_{a}\pi
\,\,\,
\hbox{ mod } 2\pi,
\end{equation}
which implies
\begin{align}
\alpha^{\,}_{a}(x)
=
l^{\,}_{a}\,\pi
-
\frac{n^{\,}_{a}\pi}{2}
\,\,\,
\hbox{ mod } 2\pi,
\quad l^{\,}_{a}=0,1,
\end{align}
\end{subequations}
since $\alpha^{\,}_{a}(x)\in[0,2\pi)$.
Imposing RS
by using the transformation rule
\eqref{eq:Sym chiral fields nu 2 c}
leads to the constraint
\begin{align}
\label{eq:H int a haldane cond odd}
h^{\,}_{a}(-x)=h^{\,}_{a}(x),
\quad
\alpha^{\,}_{a}(-x)
=
-
\alpha^{\,}_{a}(x)
\,\,\,
\hbox{ mod } 2\pi.
\end{align}
Combining TRS and RS implies that
\begin{subequations}
\label{eq: conditions on h and alpha case a}
\begin{equation}
h^{\,}_{a}(x)=h^{\,}_{a}(-x)
\label{eq: conditions on h and alpha case a a}
\end{equation}  
and
\begin{equation}
\alpha^{\,}_{a}(x)=
\left[
f^{\,}_{a}(|x|)
-
\frac{n^{\,}_{a}}{2}
\right]
\pi\,\mathrm{sgn}(x)
\,\,\,
\hbox{ mod } 2\pi,
\label{eq: conditions on h and alpha case a b}
\end{equation}
where $f^{\,}_{a}(x)$ is any function such that
\begin{align}
f^{\,}_{a}:
[0,\infty)
\to
\left\{l^{\,}_{a}\ :\ l^{\,}_{a}=0,1\right\}.
\label{eq: conditions on h and alpha case a c}
\end{align}
We note that for any even $n^{\,}_{a}$,
assuming that $f^{\,}_{a}(|x|)$ is constant,
the discontinuity at $x=0$
of $\alpha^{\,}_{a}(x)$
is an even multiple of $2\pi$
so the solution to Eqs.\
(\ref{eq: conditions on h and alpha case a b})
and
(\ref{eq: conditions on h and alpha case a c})
can be chosen independent of $x$. 
This is not the case for odd $n^{\,}_{a}$
as $n^{\,}_{a}\,\pi\,\mathrm{sgn}(x)/2\hbox{ mod }2\pi$ changes by
$\pi\hbox{ mod }2\pi$
across $x=0$.
A set of minima for the interaction
(\ref{eq:H int haldane a})
compatible with TRS and RS
that are labeled by the integers $l^{\,}_{a}$ and $n^{\,}_{a}$
are thus given by
\begin{align}
n^{\,}_{a}
\left[
\hat{\varphi}^{\,}_{\mathrm{L}}(x)
+
\hat{\varphi}^{\,}_{\mathrm{R}}(x)
\right]
+
\left(
l^{\,}_{a}\,
-
\frac{n^{\,}_{a}}{2}
\right)
\pi\,
\mathrm{sgn}(x)
=
\pi,
\label{eq: TRS and RS H int minima a}
\end{align}
\end{subequations}
where the right-hand side is defined modulo $2\pi$.
Here, to minimize the cost in kinetic energy
arising from discontinuities, we restrict discontinuities
to occur only at $x=0$ and demand that $h(x)$ vanishes smoothly
at $x=0$ if the argument of the cosine is discontinuous at $x=0$.
From now on, we only consider the cases
$n^{\,}_{a}=1$ and $n^{\,}_{a}=2$.

When $n^{\,}_{a}=1$,
the minima (\ref{eq: TRS and RS H int minima a})
simplify to
\begin{align}
\hat{\varphi}^{\,}_{\mathrm{L}}(x)
+
\hat{\varphi}^{\,}_{\mathrm{R}}(x)
=&\,
\pi
+
\left(
1/2
-
l^{\,}_{a}\,
\right)
\pi\,
\mathrm{sgn}(x)
\,\,\,
\hbox{ mod } 2\pi
\nonumber\\
=&\,
\begin{cases}
\frac{\pi}{2}\,\mathrm{sgn}(-x),&\hbox{if $l^{\,}_{a}=0$,}
\\&\\
\frac{\pi}{2}\,\mathrm{sgn}(x),&\hbox{if $l^{\,}_{a}=1$.}
\end{cases}
\label{eq: TRS and RS H int minima if na=1}
\end{align}
One verifies that
\begin{align}
\left[
\hat{\varphi}^{\,}_{\mathrm{L}}(x)
+
\hat{\varphi}^{\,}_{\mathrm{R}}(x)
\right]^{\prime}
=
\begin{cases}
\frac{\pi}{2}\,\mathrm{sgn}(-x),&\hbox{if $l^{\,}_{a}=0$,}
\\&\\
\frac{\pi}{2}\,\mathrm{sgn}(x),&\hbox{if $l^{\,}_{a}=1$,}
\end{cases}
\label{eq: TRS and RS H int minima if na=1 prime}
\end{align}
where the prime over the operators on the left-hand side
is a short-hand notation for their image under either reversal of time
or the reflection as defined by Eq.\ (\ref{eq:Sym chiral fields nu 2}).
Therefore, for a given phase profile specified by $l^{\,}_{a}$,
there exists a unique gapped ground state for
the bosonic interaction
$\widehat{H}^{\,}_{\mathrm{int}\,a}$ that is 
invariant under the action of either TRS or RS.
When $n^{\,}_{a}=1$ and the competition between the kinetic energy
and the interaction \eqref{eq:H int haldane a} results in the
opening of a spectral gap (with a midgap bound state) on the edge,
TRS and RS are neither broken explicitly nor spontaneously,
while TS is explicitly broken.
As announced below Eqs.\ (\ref{eq:imposing DIIR symmetries defs})
by making use of the bulk-edge correspondence,
the noninteracting topological classification $\mathbb{Z}$
of symmetry class DIIIR in (2+1)-dimensional spacetime reduces to the
topological classification $\mathbb{Z}^{\,}_{2}$
of symmetry class DIII when a RS compliant
breaking of TS is allowed%
~\cite{PhysRevB.88.064507,PhysRevB.88.075142},
since $\widehat{H}^{\,}_{\mathrm{int}\,a}$ with $n^{\,}_{a}=1$
is nothing but a fermionic mass term in the complex fermion
representation. The midgap states bound at the reflection symmetric 
points are protected by the actions of TRS and RS and cannot be gapped. 
Such protected ``corner" modes are nothing but the signature of a second-order 
SPT phase induced by the spatially varying mass term. Indeed,
it has been shown in Ref.\ \onlinecite{Langbehn2017}
that a two-dimensional superconductor in the symmetry class DIII
with RS but no TS 
along the boundary is an example of a second-order SPT phase~
\footnote{
The topological index belongs to the group $\mathbb{Z}^{\,}_{2}$. 
Hence, the midgap state bound by a dynamical mass supporting a domain wall
for the $\nu=4$ case can be gapped as opposed to the $\nu=2$ case.
         }.

When $n^{\,}_{a}=2$, the minima
\eqref{eq: TRS and RS H int minima a}
simplify to
\begin{subequations}
\begin{align}
2
\left[
\hat{\varphi}^{\,}_{\mathrm{L}}(x)
+
\hat{\varphi}^{\,}_{\mathrm{R}}(x)
\right]
=&\,
\pi
+
\left(
1
-
l^{\,}_{a}\,\right)
\pi\,\mathrm{sgn}(x)
\hbox{ mod } 2\pi.
\end{align}
Because
\begin{align}
\pi\,\mathrm{sgn}(x)
=
\pi
\hbox{ mod }2\pi,
\qquad
-\pi
=
\pi
\hbox{ mod }2\pi,
\end{align}
one may write 
\begin{align}
2
\left[
\hat{\varphi}^{\,}_{\mathrm{L}}(x)
+
\hat{\varphi}^{\,}_{\mathrm{R}}(x)
\right]
=
\pi\,l^{\,}_{a},
\hbox{ mod } 2\pi.
\end{align}
\end{subequations}
We conclude that
\begin{subequations}
\begin{align}
\hat{\varphi}^{\,}_{\mathrm{L}}(x)
+
\hat{\varphi}^{\,}_{\mathrm{R}}(x)
=&\,
\begin{cases}
0, &\hbox{if $l^{\,}_{a}=0$,}
\\
\pi, &\hbox{if $l^{\,}_{a}=0$,}
\\
\pi/2, &\hbox{if $l^{\,}_{a}=1$,}
\\
3\pi/2, &\hbox{if $l^{\,}_{a}=1$.}
\end{cases}
\label{eq: TRS and RS H int minima if na=2}
\end{align} 
One verifies that
\begin{align}
\left[
\hat{\varphi}^{\,}_{\mathrm{L}}(x)
+
\hat{\varphi}^{\,}_{\mathrm{R}}(x)
\right]^{\,}_{\mathrm{TRS}}
=&\,
\begin{cases}
\pi, &\hbox{if $l^{\,}_{a}=0$,}
\\
0, &\hbox{if $l^{\,}_{a}=0$,}
\\
\pi/2, &\hbox{if $l^{\,}_{a}=1$,}
\\
3\pi/2, &\hbox{if $l^{\,}_{a}=1$,}
\end{cases}
\label{eq: TRS and RS H int minima if na=2 TRS}
\end{align}
and
\begin{align}
\left[
\hat{\varphi}^{\,}_{\mathrm{L}}(x)
+
\hat{\varphi}^{\,}_{\mathrm{R}}(x)
\right]^{\,}_{\mathrm{RS}}
=&\,
\begin{cases}
0, &\hbox{if $l^{\,}_{a}=0$,}
\\
\pi, &\hbox{if $l^{\,}_{a}=0$,}
\\
3\pi/2, &\hbox{if $l^{\,}_{a}=1$,}
\\
\pi/2, &\hbox{if $l^{\,}_{a}=1$,}
\end{cases}
\label{eq: TRS and RS H int minima if na=2 RS}
\end{align}
\end{subequations}
where the subscripts TRS and RS
are short-hand notations
for the image of the minima under reversal of time and
space inversion, respectively.
There are two crucial differences between the cases
$n^{\,}_{a}=1$ and $n^{\,}_{a}=2$.
The minima
\eqref{eq: TRS and RS H int minima if na=2}
transform in a  nontrivial way under the actions of TRS and RS.
For each choice $l^{\,}_{a}$,
two minima are exchanged under the action of either
reversal of time or space inversion.
Furthermore, 
the compactness of the chiral fields and the choice
$n^{\,}_{a}=2$ conspire in such a way that they minimize the
interaction $\widehat{H}^{\,}_{\mathrm{int}\,a}$
without breaking the TS.

The cosine in the interaction $\widehat{H}^{\,}_{\mathrm{int}\,a}$
with $n^{\,}_{a}=2$ is identical to the cosine in
Hamiltonian \eqref{eq:Bos H in chiral fields nu 2 a}.
The coupling $h(x)\geq0$ breaks TS
in the interaction $\widehat{H}^{\,}_{\mathrm{int}\,a}$
when it is not a constant function of $x$, unlike
the coupling that multiplies the cosine in
Hamiltonian \eqref{eq:Bos H in chiral fields nu 2 a}.
The two choices for $l^{\,}_{a}$ in
Eq.\ (\ref{eq: TRS and RS H int minima if na=2})
correspond to fixing the overall
sign of the interaction
$\widehat{H}^{\,}_{\mathrm{int}\,a}$ with $n^{\,}_{a}=2$
when evaluated at its translation symmetric minima.
In other words, the two choices for $l^{\,}_{a}$ in
Eq.\ (\ref{eq: TRS and RS H int minima if na=2}) with $n^{\,}_{a}=2$
correspond to choosing which two translation symmetric
extrema of the cosine term are the minima. 
Furthermore, from the transformation rules
\eqref{eq: TRS and RS H int minima if na=2 TRS} and
\eqref{eq: TRS and RS H int minima if na=2 RS} we observe that
the same patterns for spontaneous symmetry-breaking patterns
as with Hamiltonian \eqref{eq:Bos H in chiral fields nu 2 a}.
When $l^{\,}_{a}=0$,
TRS is spontaneously broken, whereas RS is protected.
When $l^{\,}_{a}=1$,
RS is spontaneously broken, whereas TRS is protected.
Hence, even though the interaction \eqref{eq:H int haldane a}
breaks TS 
when $h(x)$ is not a constant function of $x$,
it shares with Hamiltonian \eqref{eq:Bos H in chiral fields nu 2 a}
the same phase diagram.

Finally, we note that the sign function that interpolates
between any two translation symmetric minima of
the interaction $\widehat{H}^{\,}_{\mathrm{int}\,a}$
also minimizes $\widehat{H}^{\,}_{\mathrm{int}\,a}$.
One verifies that this sign function respects TRS and RS but breaks
TS. Unlike the translation symmetric minima of
the interaction $\widehat{H}^{\,}_{\mathrm{int}\,a}$,
this sign function costs kinetic energy. The competition between
the kinetic and interaction terms results in a compromise by which
the singularity of the sign function is smoothed. The outcome is
a soliton that keeps TRS and RS but breaks TS. 
This soliton is a gapped excitation
that can be interpreted as a pair of helical Majorana modes
localized in the region where the soliton energy density is nonvanishing
and whose existence is protected by TRS and RS in the Majorana
representation of the  boundary theory.

\subsubsection{Symmetry constraints on Hamiltonian \eqref{eq:H int haldane b}}

PHS is again satisfied trivially by construction. 
Imposing TRS by using the transformation rule 
\eqref{eq:TRSonBFields nu 2} leads to the constraint
\begin{subequations}
\begin{align}
n^{\,}_{b}= 2 m,\qquad m\in\mathbb{Z},
\end{align}
i.e., $n^{\,}_{b}$ is an even integer.
Imposing RS by using the transformation rule \eqref{eq:RSonBFields nu 2}
leads to the pair of constraints
\begin{align}
h^{\,}_{b}(-x)=h^{\,}_{b}(x),
\qquad
\alpha^{\,}_{b}(-x)=\alpha^{\,}_{b}(x).
\end{align}
\end{subequations}
A set of minima is given by
\begin{align}
n^{\,}_{b}
\left[
\hat{\varphi}^{\,}_{\mathrm{L}}(x)
-
\hat{\varphi}^{\,}_{\mathrm{R}}(x)
\right]
+
\pi\,l^{\,}_{b}\,
=
\pi,
\,\,\,
\hbox{ mod } 2\pi,
\label{eq: TRS and RS H int minima b}
\end{align}
where $l^{\,}_{b}=0,1$. 
We only consider the case $n^{\,}_{b}=2$
and conclude that 
\begin{align}
\hat{\varphi}^{\,}_{\mathrm{L}}(x)
-
\hat{\varphi}^{\,}_{\mathrm{R}}(x)
=&\,
\begin{cases}
\pi/2, &\hbox{if $l^{\,}_{b}=0$,}
\\
3\pi/2, &\hbox{if $l^{\,}_{b}=0$,}
\\ 
0, &\hbox{if $l^{\,}_{b}=1$,}
\\
\pi, &\hbox{if $l^{\,}_{b}=1$.}
\end{cases}
\label{eq: TRS and RS H int minima if nb=2}
\end{align}
One verifies that
\begin{align}
\left[
\hat{\varphi}^{\,}_{\mathrm{L}}(x)
-
\hat{\varphi}^{\,}_{\mathrm{R}}(x)
\right]^{\,}_{\mathrm{TRS}}
=&\,
\begin{cases}
3\pi/2, &\hbox{if $l^{\,}_{b}=0$,}
\\
\pi/2, &\hbox{if $l^{\,}_{b}=0$,}
\\ 
\pi, &\hbox{if $l^{\,}_{b}=1$,}
\\
0, &\hbox{if $l^{\,}_{b}=1$.}
\end{cases}
\label{eq: TRS and RS H int minima if nb=2 TRS}
\end{align}
The four translation symmetric minima \eqref{eq: TRS and RS H int minima if nb=2}
are invariant under the action of RS. On the other hand,
under the action of TRS, two translation symmetric minima corresponding to 
each $l^{\,}_{b}$ are exchanged. Therefore, RS is always protected by
the interaction $\widehat{H}^{\,}_{\mathrm{int}\,b}$ with $n^{\,}_{b}=2$,
whereas TRS is spontaneously broken by its minima.
The argument of the cosine in
$\widehat{H}^{\,}_{\mathrm{int}\,b}$ with $n^{\,}_{b}=2$
is identical to that of the cosine in
Hamiltonian \eqref{eq:Bos H in chiral fields nu 2 b}.
Hence, both Hamiltonians obey  
the same pattern of symmetry breaking.
Finally, even though
the interaction \eqref{eq:H int haldane b} breaks 
TS when $h(x)$ is not a constant function of $x$, it shares
with Hamiltonian \eqref{eq:Bos H in chiral fields nu 2 b}
the same phase diagram (Fig.\ \ref{fig:phase diagram nu 2}).

\section{The case $\nu=1$}
\label{sec:nu 1}

For the $\nu=1$ case,
the boundary theory consists of a single helical pair of Majorana fields.
In this case, as we shall explain, it is not possible to
employ the gradient expansion method used in Sec.\ \ref{sec:nu 4}. Instead,
we proceed in two steps.  First, we establish that there are two
topological sectors in the effective bosonic theory for the boundary.
Second, we write down the dominant quartic interaction which we treat
within the mean-field approximation.

\subsection{Existence of two topological sectors}

The set (\ref{eq:ShorthandMatrix defs}) with $n=1$ has
the 4 elements ($\mathrm{X}^{\,}_{\mu}\equiv\sigma^{\,}_{\mu}$ with
$\mu=0,\cdots3$).  For $\nu=1$, there is at most $\mathrm{N}(1)=1$
interaction channel allowed by the symmetry conditions
\eqref{eq:imposing DIIR symmetries defs}.  Therefore, there is a
unique parametrization
\begin{align}
&\mathcal{H}^{(\mathrm{dyn})}_{\mathrm{bd}}(\tau,x)
\df 
\beta^{\,}_{0}\,\mathrm{i}\partial^{\,}_{x}
+
\beta^{\,}_{1}\,
\phi^{\,}(\tau,x),\\
&
\beta^{\,}_{0}\df\mathrm{X}^{\,}_{3}\equiv\sigma^{\,}_{3},
\quad
\beta^{\,}_{1}\df \mathrm{X}^{\,}_{2}\equiv \sigma^{\,}_{2},
\label{eq:def H bd nu 1}
\end{align}
of the dynamical boundary single-particle Hamiltonian.
If we impose the nonlinear constraint
\begin{equation}
\phi^{2}(\tau,x)\equiv
\bar{\phi}^{2}
\label{eq:nonlinear constraint nu 1}
\end{equation}
for some given real-valued number $\bar{\phi}$,
the target manifold is then nothing but two points $\pm1$ with the only
nonvanishing homotopy group
$\pi^{\,}_{0}(\mathsf{S}^{0})=\mathbb{Z}^{\,}_{2}$.

When the hard nonlinear constraint \eqref{eq:nonlinear constraint nu 1}
is strictly imposed, all configurations of $\phi(\tau,x)$ other than 
the constant field $\phi(\tau,x)=\pm \bar{\phi}$ must be 
discontinuous at the spacetime points where $\phi(\tau,x)$
switches between $+\bar{\phi}$ and $-\bar{\phi}$.
The gradient of $\phi(\tau,x)$ is then ill-defined
at singular points and zero everywhere else.
If we relax the condition
(\ref{eq:nonlinear constraint nu 1})
by imposing the nonlinear constraint
asymptotically,
\begin{equation}
\lim_{\tau\to\pm\infty}\phi^{2}(\tau,x)\equiv
\bar{\phi}^{2},
\label{eq:asymptotic nonlinear constraint nu 1}
\end{equation}
then smooth deformations of these singular configurations 
are admissible. However, the continuous function $\phi(\tau,x)$
then necessarily
takes the value zero along at least one time slice 
in $(1+1)$-dimensional 
space-time, which binds 
zero modes in the spectrum. This prevents employing
the gradient expansion approach outlined in Sec. \ref{sec:nu 4}
since the Pfaffian obtained by
integrating out real-valued Grassmann fields,
\begin{subequations}
\begin{align}
\label{eq:effective part func nu 1}
Z^{\,}_{\mathrm{bd}}
&\propto
\int\mathcal{D}[\phi]
\int\mathcal{D}[\chi]\,
e^{
- \int \mathrm{d}^{2}x\, 
\bar{\chi}
\left(
\mathrm{i}\gamma^{\,}_{\mu}\partial^{\,}_{\mu}	
-
\mathrm{i}\phi
\right)
\chi
}
\nonumber\\
&
\propto 
\int \mathcal{D}[\phi]\,
\mathrm{Pf}
\left[\mathrm{i}\sigma^{\,}_{2}\,\mathit{D}[\phi]\right],
\end{align}
vanishes due to zero eigenvalues of the kernel
\begin{align}
\mathit{D}\df
\mathrm{i}\gamma^{\,}_{\mu}\partial^{\,}_{\mu}	
-
\mathrm{i}\phi,
\quad
\gamma^{\,}_{0}
\df 
-
\sigma^{\,}_{2},
\quad
\gamma^{\,}_{1}
\df
\sigma^{\,}_{1},
\end{align}
\end{subequations}
where $\bar{\chi}=\chi^{\dagger} (\mathrm{i}\sigma^{\,}_{2})$.
Because the kernel $\mathrm{i}\sigma^{\,}_{2}\mathit{D}$
is skew symmetric, 
the identity
\begin{align}
\left(
\mathrm{Pf}
\left[\mathrm{i}\sigma^{\,}_{2}\,\mathit{D}[\phi]\right]
\right)^2
=
\mathrm{Det}
\left[\mathrm{i}\sigma^{\,}_{2}\,\mathit{D}[\phi]\right]
\label{eq: relation between pfaffian and determinant}
\end{align}	
holds.
Therefore,
the Pfaffian of $\mathrm{i}\sigma^{\,}_{2}\mathit{D}$,
is nothing but the square root of the functional
determinant of $\mathrm{i}\sigma^{\,}_{2}\mathit{D}$.

The idea that we shall develop below is the following.
According to Eq.\ \eqref{eq: relation between pfaffian and determinant},
computing the Pfaffian of a skew-symmetric operator is akin  
to taking the square root of a number.
Taking the square root of a real-valued number
yields two roots differing by their signs.
For any pair $\phi$ and $\phi'$, it is the relative sign
between
$
\mathrm{Pf}
\big[\mathrm{i}\sigma^{\,}_{2}\,\mathit{D}[\phi]\big]
$
and
$
\mathrm{Pf}
\big[\mathrm{i}\sigma^{\,}_{2}\,\mathit{D}[\phi']\big]
$
that fixes if $\phi$ is topologically equivalent to $\phi'$.
The background $\phi$ is topologically equivalent to $\phi'$
if
\begin{equation}
\mathrm{sgn}
\left(
\frac{
\mathrm{Pf}
\big[\mathrm{i}\sigma^{\,}_{2}\,\mathit{D}[\phi]\big]
}
{
\mathrm{Pf}
\big[\mathrm{i}\sigma^{\,}_{2}\,\mathit{D}[\phi']\big]  
}
\right)=+1.
\end{equation}
Otherwise, the background $\phi$ is not topologically equivalent to $\phi'$.
We are going to show that  there are two topological sectors in the 
effective bosonic theory, i.e., there are two disjoint 
sets of topologically inequivalent profiles of the field $\phi$.

Although the kernel $\mathrm{i}\sigma^{\,}_{2}\mathit{D}[\phi]$
is not Hermitian, the kernel
\begin{subequations}
\label{eq:equivalent kernel nu 1}
\begin{align}
&
\mathit{D}^{\prime}[\phi]\df
\begin{pmatrix} 
-\phi & +\partial \\
+\bar{\partial} & +\phi
\end{pmatrix}=
-
\mathrm{i}\sigma^{\,}_{1}\,\partial^{\,}_{x}
+
\mathrm{i}\sigma^{\,}_{2}\,\partial^{\,}_{\tau}
-
\sigma^{\,}_{3}\,\phi,
\\
&
\partial \df \partial^{\,}_{\tau}-\mathrm{i}\partial^{\,}_{x},
\quad
\bar{\partial} \df -\partial^{\,}_{\tau}-\mathrm{i}\partial^{\,}_{x},
\end{align}
\end{subequations}
(i) shares the same determinant
as $\mathrm{i}\sigma^{\,}_{2}D[\phi]$
and (ii) is Hermitian. It follows that
the eigenvalues of $D^{\prime}[\phi]$
are real valued. 
Moreover, the kernel $\mathit{D}^{\prime}[\phi]$ obeys the
Bogoliubov-de Gennes condition and, hence, the
nonvanishing real-valued eigenvalues of $\mathit{D}^{\prime}[\phi]$
come in pairs of opposite signs. We shall assume that all eigenvalues
of $\mathit{D}^{\prime}[\phi]$ are nonvanishing. The label $\iota$
enumerates all pairs of eigenvalues
$\pm|\lambda^{\prime}_{\iota}|\in\mathbb{R}\setminus\{0\}$
of $\mathit{D}^{\prime}[\phi]$.
We then have the definition
\begin{equation}
\begin{aligned}[b]
\mathrm{Pf}\left[\mathrm{i}\sigma^{\,}_{2}\,\mathit{D}[\phi]\right]
& \df
\prod_{\iota}
|\lambda^{\prime}_{\iota}|
\end{aligned}	
\label{eq:Pfaffian as Eigenvalues nu 1}
\end{equation}
that consists of choosing all the positive representatives of the
pairs of nonvanishing eigenvalues.
The question that immediately arises is if this definition can be done
consistently over the entire target space of $\phi$. If the answer
to this question is positive, then the target space is topologically trivial.
Otherwise, it is not.

Our goal is to show that there are two distinct
topological sectors as discussed above.
To this end, we shall choose an arbitrary profile
$\phi(\tau,x)$ that obeys the boundary conditions
\eqref{eq:asymptotic nonlinear constraint nu 1}
and prove the identities
\begin{subequations}
\label{eq:spectral flow relations nu 1}
\begin{align}
\label{eq:spectral flow relations a nu 1}
\mathrm{sgn}
\left(
\frac{
\mathrm{Pf}
\big[\mathrm{i}\sigma^{\,}_{2}\,\mathit{D}[\phi]\big]
}
{
\mathrm{Pf}
\big[\mathrm{i}\sigma^{\,}_{2}\,\mathit{D}[\bar{\phi}]\big]  
}
\right)
=
-
\mathrm{sgn}
\left(
\frac{
\mathrm{Pf}
\big[\mathrm{i}\sigma^{\,}_{2}\,\mathit{D}[\phi]\big]
}
{
\mathrm{Pf}
\big[\mathrm{i}\sigma^{\,}_{2}\,\mathit{D}[-\bar{\phi}]\big]  
}
\right),
\end{align}
and
\begin{align}
\label{eq:spectral flow relations b nu 1}
\mathrm{sgn}
\left(
\frac{
\mathrm{Pf}
\big[\mathrm{i}\sigma^{\,}_{2}\,\mathit{D}[\phi]\big]
}
{
\mathrm{Pf}
\big[\mathrm{i}\sigma^{\,}_{2}\,\mathit{D}[-\phi]\big]  
}
\right)
=
-1.
\end{align}
\end{subequations}
Two comments are in order before we prove Eqs.\ 
\eqref{eq:spectral flow relations nu 1}.
Equation \eqref{eq:spectral flow relations a nu 1} 
implies that profile $\phi$ is topologically equivalent to 
either one of the two constant profiles $\pm\bar{\phi}$.
In other words, there exist exactly two topological sectors
with representative profiles $+\bar{\phi}$ and $-\bar{\phi}$
as measured by Eq.\ \eqref{eq:spectral flow relations a nu 1}.
Equation \eqref{eq:spectral flow relations b nu 1}
implies that the profiles $\phi$ and $-\phi$ belong 
to distinct topological sectors,
a fact that originates from a $\mathbb{Z}^{\,}_{2}$
global anomaly~\cite{WITTEN1982324,PhysRevLett.99.116601}. 
Indeed, the transformation 
\begin{subequations}
\label{eq:anomaly signam3 nu 1}
\begin{align}
\label{eq:sigma3 transf nu 1}
\chi = \sigma^{\,}_{3} \chi',
\quad
\phi = -\phi',
\end{align}
leaves the Lagrangian
\begin{align}
\bar{\chi}
\left(
\mathrm{i}\gamma^{\,}_{\mu}\partial^{\,}_{\mu}	
-
\mathrm{i}\phi
\right)
\chi
=
\bar{\chi}'
\left(
\mathrm{i}\gamma^{\,}_{\mu}\partial^{\,}_{\mu}	
-
\mathrm{i}\phi'
\right)
\chi'
\end{align}
invariant, while
the partition function \eqref{eq:effective part func nu 1} 
changes according to 
\begin{align}
Z^{\,}_{\mathrm{bd}}
&\propto
\int 
\mathcal{D}[\phi']
\mathcal{D}[\chi']
\mathcal{J}\left[\sigma^{\,}_{3}\right]
e^{
- \int \mathrm{d}^{2}x\, 
\bar{\chi'}
\left(
\mathrm{i}\gamma^{\,}_{\mu}\partial^{\,}_{\mu}	
-
\mathrm{i}\phi'
\right)
\chi'}
\nonumber\\
&
\propto
\int 
\mathcal{D}[\phi']
\mathcal{J}\left[\sigma^{\,}_{3}\right]
\mathrm{Pf}
\big[\mathrm{i}\sigma^{\,}_{2}\,\mathit{D}[\phi']\big]
\nonumber\\
&
\propto
\int 
\mathcal{D}[\phi]
\mathcal{J}\left[\sigma^{\,}_{3}\right]
\mathrm{Pf}
\big[\mathrm{i}\sigma^{\,}_{2}\,\mathit{D}[-\phi]\big].
\end{align}
On the one hand, to reach the right-hand 
side of the second line, we allowed for
a possibly nontrivial Jacobian $\mathcal{J}[\sigma^{\,}_{3}]$
associated with the transformation $\chi = \sigma^{\,}_{3} \chi'$.
On the other hand, to reach the third line, we assumed that 
the Jacobian associated with the transformation $\phi = -\phi'$
is unity. 
Equation \eqref{eq:spectral flow relations b nu 1} then implies that
$\mathcal{J}\left[\sigma^{\,}_{3}\right] = -1$, which is the 
precise definition of a $\mathbb{Z}^{\,}_{2}$
global anomaly, namely the symmetry of the Lagrangian that is not 
respected by the measure. 
\end{subequations}

{\it Proof of Eqs.\ \eqref{eq:spectral flow relations nu 1}.}
We now prove Eqs.\ \eqref{eq:spectral flow relations nu 1}.
To examine whether two profiles
$\phi^{\,}_{\mathrm{i}}(\tau,x)$
and $\phi^{\,}_{\mathrm{f}}(\tau,x)$ 
are topologically equivalent, 
we introduce a parameter $t\in [0,1]$ 
and define a continuous function $\phi^{\,}_{t}(\tau,x)$
such that
\begin{subequations}
\label{eq:SpectralFlowEq nu 1}
\begin{equation}
\phi^{\,}_{t=0}(\tau,x) =\phi^{\,}_{\mathrm{i}}(\tau,x),
\quad
\phi^{\,}_{t=1}(\tau,x) =
\phi^{\,}_{\mathrm{f}}(\tau,x).
\end{equation}
We choose the linear interpolation 
\begin{align}
\phi^{\,}_{t}(\tau,x)
\df
(1-t)\,\phi^{\,}_{\mathrm{i}}(\tau,x)
+
t\,
\phi^{\,}_{\mathrm{f}}(\tau,x).
\end{align}
\end{subequations}
We impose periodic boundary conditions in both $\tau$
and $x$,
\begin{align}
\label{eq:periodic boound cond nu 1}
\phi(\tau,x+L^{\,}_{x}) = \phi(\tau,x),
\quad
\phi(\tau+L^{\,}_{\tau},x) = \phi(\tau,x).
\end{align}
Hence, interpolation
\eqref{eq:SpectralFlowEq nu 1} also satisfies
these boundary conditions.
Boundary conditions \eqref{eq:periodic boound cond nu 1}
describe a compact space-time
($\mathsf{S}^{1}\times \mathsf{S}^{1}=\mathsf{T}^{2}$). 
It follows that the spectrum of the kernel
$\mathit{D}^{\prime}[\phi^{\,}_{t}]$ defined in
Eq.\ \eqref{eq:equivalent kernel nu 1}
is discrete.
If one calculates the flow of eigenvalues 
$\lambda^{\prime}_{t,\iota}$ of the kernel
$\mathit{D}^{\prime}[\phi^{\,}_{t}]$ as a function of $t$,
whenever there is a gap closing, i.e., 
at least one of the $\lambda^{\prime}_{t,\iota}$
is $0$,  there is a $\pi$ phase change in the Pfaffian.
Thus, an odd number of gap closings 
during the evolution from $t=0$ to $t=1$
means that the initial and final profiles
belong to different topological sectors. 
We will prove Eqs.\ \eqref{eq:spectral flow relations nu 1}
by assuming that the number of gap closings is independent
of the choice of the interpolation scheme, 
without calculating the actual number of gap closings explicitly.

We first examine a special case of Eq.\ 
\eqref{eq:spectral flow relations a nu 1} for which 
$\phi(\tau,x)=+\bar{\phi}$. 
Consider the linear interpolation
\begin{align}
\phi^{+,-}_{t}
\df
(1-t)\bar{\phi}
+
t (-\bar{\phi})
=
(1-2t)\bar{\phi}.
\end{align}
For any $t\neq 1/2$, $\phi^{+,-}_{t}$ 
contributes to the Kernel $\mathit{D}^{\prime}[\phi^{+,-}_{t\neq1/2}]$
as a constant nonvanishing mass term. Hence, the 
spectrum is gapped. This gap closes only at $t=1/2$,
in which case the kernel $\mathit{D}^{\prime}[\phi^{+,-}_{t=1/2}]$
is that of a free Majorana fermion. There exists only a single 
pair of zero eigenvalues that are labeled by reciprocal 
vector $(\omega,k)=(0,0)$. 
Therefore, we find that there is a single crossing
between negative and positive eigenvalues of 
$\mathit{D}^{\prime}[\phi^{+,-}_{t}]$ at $t=1/2$.
It follows that in the special case $\phi(\tau,x)=+\bar{\phi}$,
Eq.\ \eqref{eq:spectral flow relations a nu 1} holds. 
For any profile $\phi(\tau,x)$, the manipulation
\begin{align}
\label{eq:manipulation sign of pfaffian nu 1}
\mathrm{sgn}
\left(
\frac{
\mathrm{Pf}
\big[\mathrm{i}\sigma^{\,}_{2}\,\mathit{D}[\phi]\big]
}
{
\mathrm{Pf}
\big[\mathrm{i}\sigma^{\,}_{2}\,\mathit{D}[\bar{\phi}]\big]  
}
\right)
&=
\mathrm{sgn}
\left(
\frac{
\mathrm{Pf}
\big[\mathrm{i}\sigma^{\,}_{2}\,\mathit{D}[\phi]\big]
}
{
\mathrm{Pf}
\big[\mathrm{i}\sigma^{\,}_{2}\,\mathit{D}[-\bar{\phi}]\big]  
}
\right)
\nonumber\\
&
\times
\mathrm{sgn}
\left(
\frac{
\mathrm{Pf}
\big[\mathrm{i}\sigma^{\,}_{2}\,\mathit{D}[-\bar{\phi}]\big]
}
{
\mathrm{Pf}
\big[\mathrm{i}\sigma^{\,}_{2}\,\mathit{D}[\bar{\phi}]\big]  
}
\right)
\end{align}
then implies Eq.\ \eqref{eq:spectral flow relations a nu 1}.
Observe that 
identity \eqref{eq:manipulation sign of pfaffian nu 1}
is nothing but the interpolation
\begin{align}
\Phi^{+,-}_{t}
\df
\begin{cases}
(1-2t)\phi(\tau,x)
-
2t\,\bar{\phi},
&
\text{if $0\leq t < \frac{1}{2}$},
\\
(2t-2)\bar{\phi}
+
(2t-1)\bar{\phi},
&
\text{if $\frac{1}{2} \leq t \leq 1$}.
\end{cases}
\end{align}

To show Eq.\ \eqref{eq:spectral flow relations b nu 1},
we note that for any $\phi(\tau,x)$,
\begin{align}
\sigma^{\,}_{2}\,\mathsf{K}\,
D^{\prime}[\phi]\,
\mathsf{K}\,
\sigma^{\,}_{2}
=
D^{\prime}[-\phi].
\end{align}
Hence, $D^{\prime}[\phi]$ and $D^{\prime}[-\phi]$
share the same eigenvalue spectrum.
This implies that
for the two interpolations 
\begin{subequations}
\begin{align}
&
\Phi^{+}_{t}
\df
(1-t)\,\bar{\phi}
+
t\, \phi(\tau,x),
\\
&
\Phi^{-}_{t}
\df
(1-t)\,(-\bar{\phi})
+
t\, (-\phi(\tau,x))
=
-\Phi^{+}_{t},
\end{align}
$D^{\prime}[\Phi^{+}_{t}]$ and $D^{\prime}[\Phi^{-}_{t}]$
also share the same eigenvalue spectrum.
Therefore, one can then show that
\begin{align}
\mathrm{sgn}
\left(
\frac{
\mathrm{Pf}
\big[\mathrm{i}\sigma^{\,}_{2}\,\mathit{D}[\phi]\big]
}
{
\mathrm{Pf}
\big[\mathrm{i}\sigma^{\,}_{2}\,\mathit{D}[\bar{\phi}]\big]  
}
\right)
=
\mathrm{sgn}
\left(
\frac{
\mathrm{Pf}
\big[\mathrm{i}\sigma^{\,}_{2}\,\mathit{D}[-\phi]\big]
}
{
\mathrm{Pf}
\big[\mathrm{i}\sigma^{\,}_{2}\,\mathit{D}[-\bar{\phi}]\big]  
}
\right),
\end{align}
which after rearrangement gives
\begin{align}
\mathrm{sgn}
\left(
\frac{
\mathrm{Pf}
\big[\mathrm{i}\sigma^{\,}_{2}\,\mathit{D}[\phi]\big]
}
{
\mathrm{Pf}
\big[\mathrm{i}\sigma^{\,}_{2}\,\mathit{D}[-\phi]\big]  
}
\right)
&=
\mathrm{sgn}
\left(
\frac{
\mathrm{Pf}
\big[\mathrm{i}\sigma^{\,}_{2}\,\mathit{D}[\bar{\phi}]\big]
}
{
\mathrm{Pf}
\big[\mathrm{i}\sigma^{\,}_{2}\,\mathit{D}[-\bar{\phi}]\big]  
}
\right)
\nonumber\\
&=
-1.
\end{align}
\end{subequations}
Any profile $\phi(\tau,x)$ is topologically inequivalent 
to $-\phi( \tau,x)$, as claimed in 
Eq.\ \eqref{eq:spectral flow relations b nu 1}.

\subsection{Mean-field treatment of the interaction}

To complement the discussion in the previous subsection,
we integrate over the bosonic field $\phi$ in action 
\eqref{eq:effective part func nu 1}
and derive the effective action for the Majorana fields 
$\hat{\chi}^{\,}_{\mathrm{L}}$ and $\hat{\chi}^{\,}_{\mathrm{R}}$.
The single interaction term has the form
$\hat{\chi}^{\,}_{\mathrm{L}}(x)\,
\hat{\chi}^{\,}_{\mathrm{R}}(x)\,
\hat{\chi}^{\,}_{\mathrm{L}}(x+\epsilon)\,
\hat{\chi}^{\,}_{\mathrm{R}}(x+\epsilon)$
where $\epsilon$ is a short-distance cutoff that implements point splitting.
For weak coupling strength this interaction term is irrelevant and the 
boundary remains gapless. In the limit of a strong interaction
strength, a gap opens in the spectrum \cite{Aasen2020,Chou2021}.
At the mean-field level, this gap corresponds to the bilinear 
$\mathrm{i}\hat{\chi}^{\,}_{\mathrm{L}}\,\hat{\chi}^{\,}_{\mathrm{R}}$
acquiring a nonvanishing expectation value. 
This is equivalent to replacing the dynamical field
$\phi(\tau,x)$ in action \eqref{eq:effective part func nu 1} by the constant
profiles $\pm \bar{\phi}$. Inserting the mean-field solution for the field 
$\phi(\tau,x)$ explicitly breaks the TRS since the term 
$\pm\mathrm{i}\,\bar{\phi}\,
\hat{\chi}^{\,}_{\mathrm{L}}\,\hat{\chi}^{\,}_{\mathrm{R}}$
is odd under the transformation \eqref{eq:SymmetryOperators b defs}.
Gapping the boundary is only possible by spontaneously breaking 
TRS. 

\section{Conclusion}
\label{sec:Conclusion}

We have studied by nonperturbative means
the stability of a two-dimensional crystalline topological superconductor
in symmetry class DIIIR
when perturbed by symmetry-preserving
quartic contact interactions. 
Building on the fact that eight copies of helical pairs of edge
modes are gapped by such interactions
without symmetry breaking,
we investigated the stability of $\nu=1,2,4$
copies of the helical pairs of edge theories 
in order to understand how these
cases remain stable. 
For $\nu=4$ copies of edge modes, we identified
four interacting channels and presented an analytical
derivation of the low-energy effective action,
which is a NLSM model supplemented by a WZ term.
In $(1+1)$ dimensions, this action flows to that of
a gapless theory.
Hence, the interacting theory remains gapless.
We then employed bosonization methods to study 
interactions between $\nu=2$ copies of helical pairs of edge modes.
We found that there is a regime in coupling space
for which interactions become relevant,
but always at the cost of the spontaneous breaking of one
of the two protecting symmetries, provided we impose
translation symmetry on the edge.
For the final case of a single helical pair of edge modes, 
although we were not able to bosonize the fermionic theory explicitly,
we showed that there exist two topological sectors
and a $\mathbb{Z}^{\,}_{2}$ global anomaly.
We instead analyzed the stability of the noninteracting
edge states by using their Majorana representation and showed that the
boundary can only be gapped at the cost of spontaneously breaking 
the TRS.

In two-dimensional space, the symmetry class BDIR
corresponds to a TRS crystalline superconductor for which
the operation of time reversal squares to $+1$.
Its noninteracting topological classification $\mathbb{Z}$ becomes
the classification $\mathbb{Z}^{\,}_{8}$ in the presence of
symmetry preserving contact quartic interactions\,
\cite{PhysRevB.95.195108}.
Our approach would also apply to this case.
The stability analysis of two-dimensional crystalline insulators with
noninteracting topological classification $\mathbb{Z}$
can always be dealt with using Abelian bosonization techniques in combination
with the Haldane criterion. However, Abelian bosonization techniques
are not applicable to three-dimensional space. Instead, one relies on
functional bosonization techniques based
on the gradient expansion or on conjectured dualities.
Detailed stability analysis of some two- and three-dimensional
topological crystalline insulators can be found in
Refs.\ \onlinecite{Yoshida2015,Isobe2015,Hsieh2016}.
   
\section*{Acknowledgments}

\"O.M.A. was supported by the Swiss National Science Foundation (SNSF)
under Grant No.\ 200021 184637.  J.-H.C.\ was supported by the SNSF
through Grants No.\ 2000021 153648 and P2EZP2-184306.  S.R. is
supported by the National Science Foundation under Award No.
DMR-2001181, and by a Simons Investigator Grant from the Simons
Foundation (Award No. 566116).  A.F. was supported by JSPS KAKENHI
(Grant No. 19K03680) and JST CREST (Grant No. JPMJCR19T2).

\bibliography{References}
\end{document}